\documentclass[%
twocolumn,
longbibliography,
amsmath,amssymb,
aps,prl,
floatfix,
]{revtex4-1}

\usepackage{graphicx}
\usepackage{dcolumn}
\usepackage{bm}
\usepackage{color}

\usepackage{xcolor}
\newcommand{\CS}[1]{\textcolor{violet}{{} }}
\newcommand{\rev}[1]{\textcolor{black}{#1}}

\begin{document}

\title{Exploiting Aharonov-Bohm oscillations to probe Klein tunneling \\ in tunable pn-junctions in graphene}

\author
{J. Dauber$^{1,2}$, K. J. A. Reijnders$^{3}$, L. Banszerus$^{1,2}$, A. Epping $^{1,2}$, K. Watanabe$^4$, \\ 
T. Taniguchi$^5$, M. I. Katsnelson$^{3}$, F. Hassler$^{6}$ and C. Stampfer$^{1,2,*}$
\\
\normalsize{$^1$JARA-FIT and 2nd Institute of Physics, RWTH Aachen University, 52074 Aachen, Germany, EU}\\
\normalsize{$^2$Peter Gr\"unberg Institute (PGI-9), Forschungszentrum J\"ulich, 52425 J\"ulich, Germany, EU}\\
\normalsize{$^3$Radboud University, Institute for Molecules and Materials, 6525AJ Nijmegen, The Netherlands, EU}\\
\normalsize{$^4$Research Center for Functional Materials, National Institute for Materials Science, 1-1 Namiki Tsukuba, Ibaraki 305-0044, Japan}\\
\normalsize{$^5$International Center for Materials Nanoarchitectonics, 
National Institute for Materials Science, 1-1 Namiki   Tsukuba Ibaraki
305-0044, Japan}\\
\normalsize{$^6$JARA-Institute for Quantum Information, RWTH Aachen University, 52056 Aachen, Germany, EU}
\\
\normalsize{$^{*}$Corresponding author; E-mail: stampfer@physik.rwth-aachen.de}
}

\date{\today}

\begin{abstract}
One of the unique features of graphene is that the Fermi wavelength of its charge carriers can be tuned electrostatically over a  wide range. 
This allows in principle to tune  the transparency of a pn-junction electrostatically, as this depends on the ratio between the physical extension of the junction and the electron wavelength, i.e. on the effective width of the junction itself. However, this simple idea ---  which would allow to switch smoothly between a Veselago lens and a Klein-collimator --- has proved to be  difficult to  demonstrate  experimentally because of the limited amount of independently-tunable parameters available in most setups.  In this work, we present transport measurements in a quasi-ballistic Aharonov-Bohm graphene ring with gate tunable pn-junctions in one arm, and show that the interference patterns provide unambiguous information on the Klein tunneling efficiency
and on the junctions effective width. \rev{We find a gate-controlled transparency of the pn-junctions ranging from 35--100\%}.  
Our results are in excellent agreement with \rev{a semiclassical description}.
\end{abstract}

\maketitle

Graphene represents an attractive platform to study coherent electron optics in two dimensions~\cite{Boggild2017Jun}. The Dirac-fermion nature of its charge carriers allows to tune the electron wavelength over a wide range, and enables negative refraction at an interface between a hole and an electron doped region, forming a pn-junction~\cite{Cheianov2007Mar,Lee2015Sep,Chen2016}. This phenomenon, along with a low-temperature mean free-path of more than 25~$\mu$m~\cite{Banszerus2016Feb} and a phase coherence length of several micrometers, makes graphene an ideal material for quantum electron optic devices, such as Fabry-P\'erot interferometers~\cite{Young2009Feb,Rickhaus2013Aug}, ballistic switches~\cite{Sajjad2013Nov,Wang2019Apr,Graef2019Jun} as well as  Klein tunneling transistors~\cite{Wilmart2014Apr} and tunable wave guides~\cite{Williams2011Feb,Rickhaus2015Sep}.
All these devices are based on pn-junctions and their operation depends crucially on the effective width $d_\text{pn} = w_\text{pn}/ \lambda_\text{F}$ of the \rev{pn-}junction itself, which  measures the extent (width) $w_\text{pn}$ of the junction with respect to the Fermi length $\lambda_\text{F}$. 
At sharp pn-interfaces ($d_\text{pn} \lesssim 1$), Klein-like tunneling~\cite{Klein1929,Katsnelson2006} leads to a refocusing of the divergent electron beams similar to a Veselago lens~\cite{Cheianov2007Mar,Reijnders17b,Reijnders17,Brun2019Jul}, and results in a high overall transparency. 
For wide pn-junctions ($d_\text{pn} > 1$), only electrons with near-perpendicular incidence are allowed to pass, thus forming a so-called Klein collimator~\cite{Cheianov2006, Libisch2017Feb}. 
The latter is important for ballistic switches and electron wave guides~\cite{Sajjad2013Nov,Wang2019Apr,Wilmart2014Apr,Williams2011Feb,Rickhaus2015Sep}, but exhibits limited average transmission through the junction. As  the Fermi wavelength $\lambda_\text{F} \propto |n|^{-1/2}$ can be controlled by tuning the carrier density $n$, the effective width $d_\text{pn}$ can be in principle tuned in-situ by electrostatic gates. 
However, the experimental demonstration of \rev{the tunable effective width of a} pn-junction has proved difficult to realize because of severe limitations on the number of independently tunable parameters provided by most detection schemes. Furthermore, it is typically difficult to distinguish effects originating from tunable pn-junctions from other mesoscopic effects, such as boundary or edge scattering, and  contact resistances. 

\begin{figure*}[tb]\centering 
\includegraphics[draft=false,keepaspectratio=true,clip,%
                   width=1\linewidth]%
                   {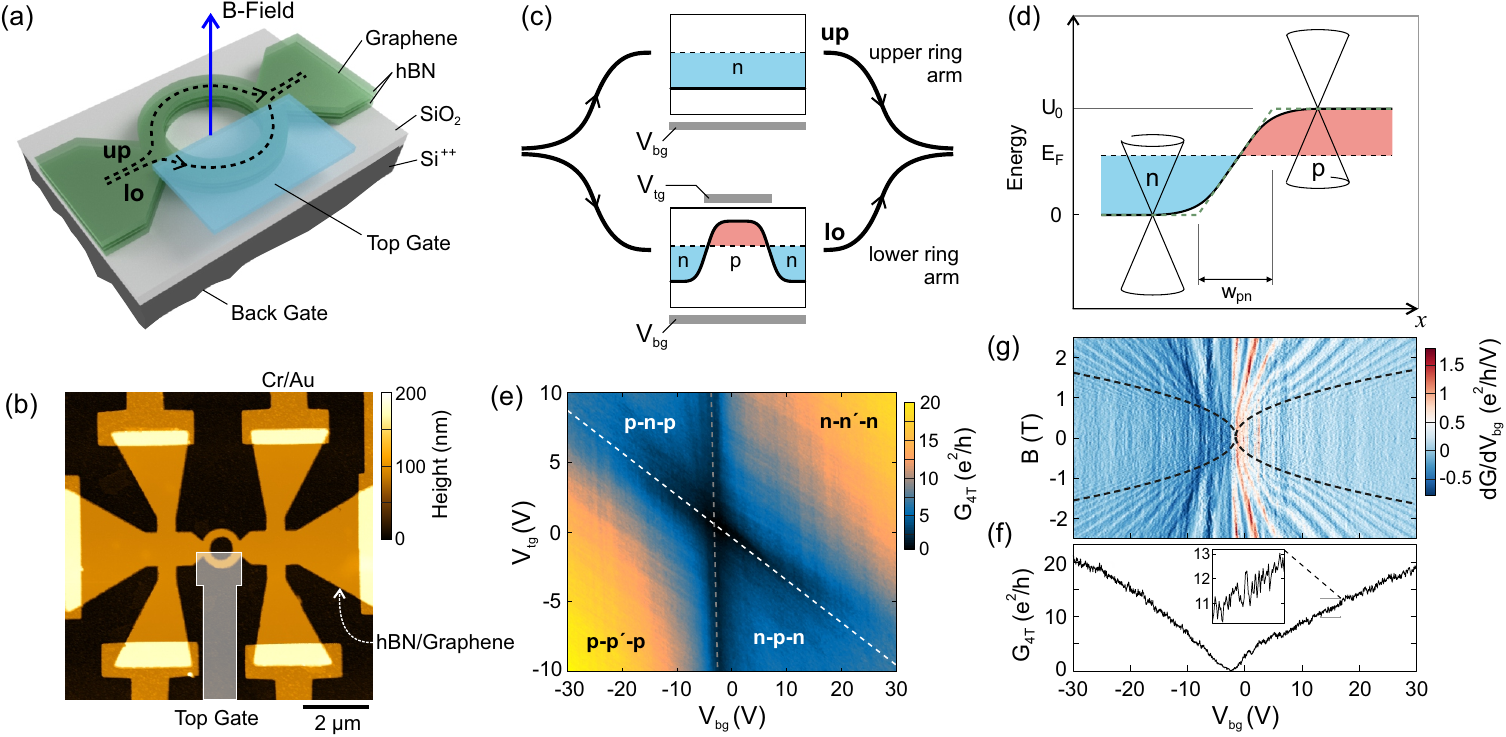}                   
\caption{\textbf{(a)} 
Schematic illustration of the graphene-based ring device highlighting the electron/hole trajectories along the upper and lower arm of the ring. 
\textbf{(b)} 
Scanning force microscope image of the investigated device before placing the top-most hBN layer and the top gate electrode (more details in Supp. Material). 
The location of the top gate is highlighted by the shaded area.
\textbf{(c)} Illustration of the electrostatic tunability of the ring device allowing to implement a npn-junction (two pn-junctions in series) in the lower ring arm. 
\textbf{(d)} Schematics exhibiting the energy scales and the geometric width $w_{\text{pn}}$ of 
a pn-junction.
\textbf{(e)} Color plot of the four terminal conductance $G_{\text{4T}}$ as function of $V_{\text{bg}}$ and $V_{\text{tg}}$.
The dashed lines and labels highlight the different configurations with and without pn-junctions in the lower arm of the ring.
\textbf{(f)} Four terminal conductance along $V_{\text{tg}}=0.25 \times V_{\text{bg}}+0.625\,\text{V}$.
\rev{\textbf{(g)} Derivative of the two-terminal conductance with respect to gate voltage $dG/dV_{\text{bg}}$
as function of $V_{\text{bg}}$ (with varying $V_{\text{tg}}$, see panel f) and $B$-field. Dashed lines highlight constant cyclotron radius of $r_\mathrm{C} = w/2$.}
} 
\label{figure1}
\end{figure*}

Here, we overcome these limitations by employing an Aharonov-Bohm interferometer in a ballistic graphene ring as a probe for transmission (see Fig.~1a). 
Using the amplitude of the Aharonov-Bohm interference oscillations to extract the transmission coefficient~\cite{Schelter2010,Schelter2012Aug}, we are able to demonstrate the tunability of the effective width of the pn-junction, realizing both large and small values of $d_\text{pn}$.  
We show that for a small effective pn-junction width a fully symmetric response is obtained for pn- and n$'$n-junctions, indicating a loss-less tuning of the optical refractive index. Our result not only provides the long missing experimental evidence for the large tunability of the effective width of pn-junctions, but open the door to  the realization of more complex coherent electron optic devices based on graphene with pn-junctions. We thus present the realization of a crucial building block for future applications of Dirac fermion optic devices.

The Aharonov-Bohm (AB) effect~\cite{Aharonov1959,Tonomura1982May} is sensitive to the transmission properties of the coherently propagating partial electron waves, and it is therefore suitable for probing the transmission properties of a pn-junction inserted into the ring~\cite{Schelter2010}.
Although the AB effect has been observed in various solid-state systems~\cite{Webb1985Jun,Timp1987Jun,Gao1994Dec,Appenzeller1995Feb}, AB rings in graphene have either been in the diffusive regime~\cite{Russo2008,Huefner2010,Yoo2010,Smirnov2012,Nam2012Dec} or lacked the necessary geometry to exploit ballistic and phase-coherent transport through tunable pn- junctions~\cite{Dauber2017Nov}.

Our device is based on graphene grown by chemical vapor deposition (CVD), which is encapsulated in hexagonal boron nitride (hBN) by dry van der Waals assembly~\cite{Banszerus2015}. The hBN/graphene/hBN heterostructure is structured by dry etching into a ring shaped device (Fig.~\ref{figure1}b) with a design similar to previous works on III-V semiconductors~\cite{Grbic2007}. The design is optimized such that electron trajectories with many different angles may enter the ring (see Figs.~\ref{figure1}a,b). The leads have a width of 1~$\mu$m, the ring has a mean radius of $\overline{r} = 500$~nm, and the ring arms have a width of \rev{$w=200$~nm}. After placing metal contacts for four-terminal measurements, an additional hBN crystal, and a structured metallic top gate are placed on top, where the gate electrode covers only the lower ring arm (see shaded area in Fig.~\ref{figure1}b and Supp. Material), \rev{similar to the device studied in Ref.~\cite{Smirnov2012}}. 
The global back gate and the local top gate allow separate control over the carrier densities in the two ring arms as illustrated in Fig.~\ref{figure1}c. Most importantly, the independent control of the back gate voltage $V_{\text{bg}}$ and the top gate voltage $V_{\text{tg}}$ enables the formation of two pn-junctions in series along the lower ring arm. As the carrier densities in the lead region ($n \approx \alpha_{\text{bg}} V_{\text{bg}}$) and 
in the lower ring arm ($n_{\text{lo}} \approx n + \alpha_{\text{tg}} V_{\text{tg}}$) are tuned independently, 
both the Fermi energy $E_\text{F} \approx n \hbar v_\text{F} \sqrt{\pi/|n|}$ and the barrier height $U_0 \approx E_\text{F} + n_{\text{lo}} \hbar v_\text{F} \sqrt{\pi/|n_{\text{lo}}|}$ (in the lower arm) are controlled separately (see Figs.~\ref{figure1}c,d).
Here, $v_\text{F} \approx 10^6$~m/s is the Fermi velocity and $\alpha_{\text{bg}} \approx 6.4 \times 10^{10}$~cm$^{-2}$V$^{-1}$ and $\alpha_{\text{tg}} \approx 2.6 \times 10^{11}$~cm$^{-2}$V$^{-1}$ are the capacitive back and top gate lever arms, respectively (for more details see Supp. Material).
As the total thickness of the \rev{two} hBN crystals separating the top gate from the graphene is around 60~nm, we estimate the geometric pn-junction width to be also around $w_{\text{pn}} \approx 60$~nm~\cite{Chen2016}.

\begin{figure}[tb]\centering 
\includegraphics[draft=false,keepaspectratio=true,clip,%
                   width=0.82\linewidth]%
                   {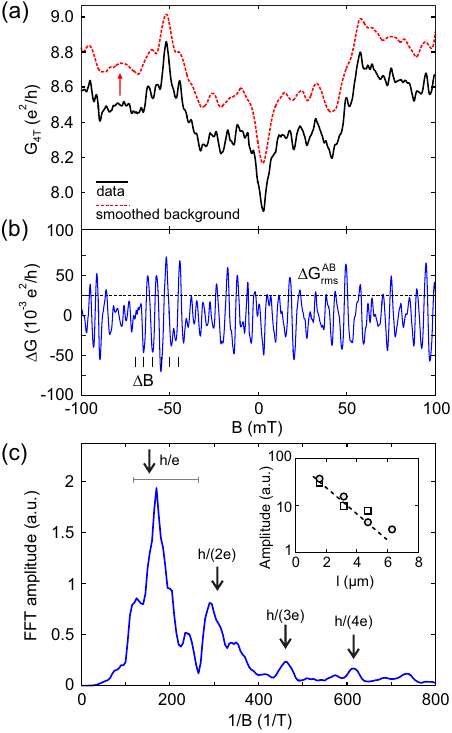}                   
\caption{Aharonov-Bohm oscillations. \textbf{(a)} Four-terminal magnetoconductance measured in the nn$'$n regime at $V_{\text{bg}}=10\,\text{V}$ and $V_{\text{tg}}=4.5\,\text{V}$. The dashed line (plotted offset) shows the smoothed data that is used for the background subtraction. \textbf{(b)} Background subtracted conductance of data shown in panel (a). The dashed line depicts the RMS value of $\Delta G$ that is used to measure the amplitude of the AB oscillations. The vertical lines indicate the AB periodicity $\Delta B$. \textbf{(c)} Corresponding Fourier transform, where the arrows mark the observed fundamental mode of the AB oscillations and their expected higher harmonics ($h/(m e)$; see labels). The inset shows the scaled amplitude (circles) of the FFT peaks as function of $l=m \pi \overline{r}$ ($m=1,2,...$). From the exponential decay $\propto \exp(-l/l_{\phi})$, we estimate $l_{\phi} \approx 1.5\,\mu$m (dashed line). The squares represent another data set at different gate voltages.
} 
\label{figure2}
\end{figure}

\begin{figure}[tb]\centering 
\includegraphics[draft=false,keepaspectratio=true,clip,%
                   width=0.9\linewidth]%
                   {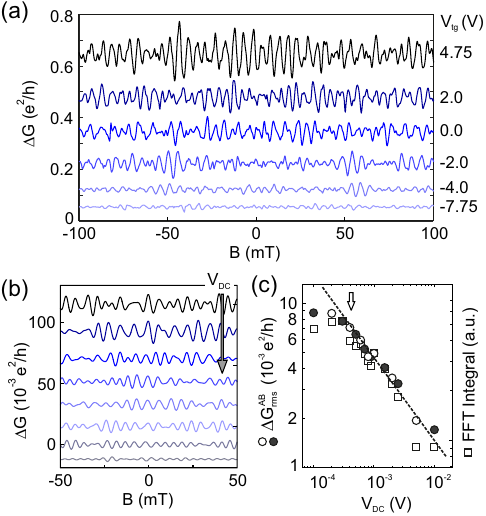}                   
\caption{
\rev{\textbf{(a)} AB oscillations in the background subtracted conductance for different $V_{\text{tg}}$ (see labels) at a fixed  $V_{\text{bg}} = 30$~V, highlighting the $V_{\text{tg}}$ dependence of the AB amplitudes. The traces are vertically offset for clarity (see dashed lines).
\textbf{(b)} AB oscillations at fixed gate voltages ($V_{\text{bg}} = -6$~V and $V_{\text{tg}} = -2.5$~V) but varying DC bias, $V_{\text{DC}}$ in the range of $0.1$ to $10$~mV (see filled data points in panel c). The traces are vertically offset for clarity (see dashed lines).
\textbf{(c)} $\Delta G_{\text{rms}}^{\text{AB}}$ as a function of $V_{\text{DC}}$ (circles). The squares show the integral of the FFT amplitude of the $h/e$ mode (for more details see supplemental Fig. S7).
The solid line represent the $1/\sqrt{V_{\text{DC}}}$ dependency.
}
} 
\label{figure3new}
\end{figure}

\begin{figure*}[tb]\centering 
\includegraphics[draft=false,keepaspectratio=true,clip,%
                   width=1.0\linewidth]%
                   {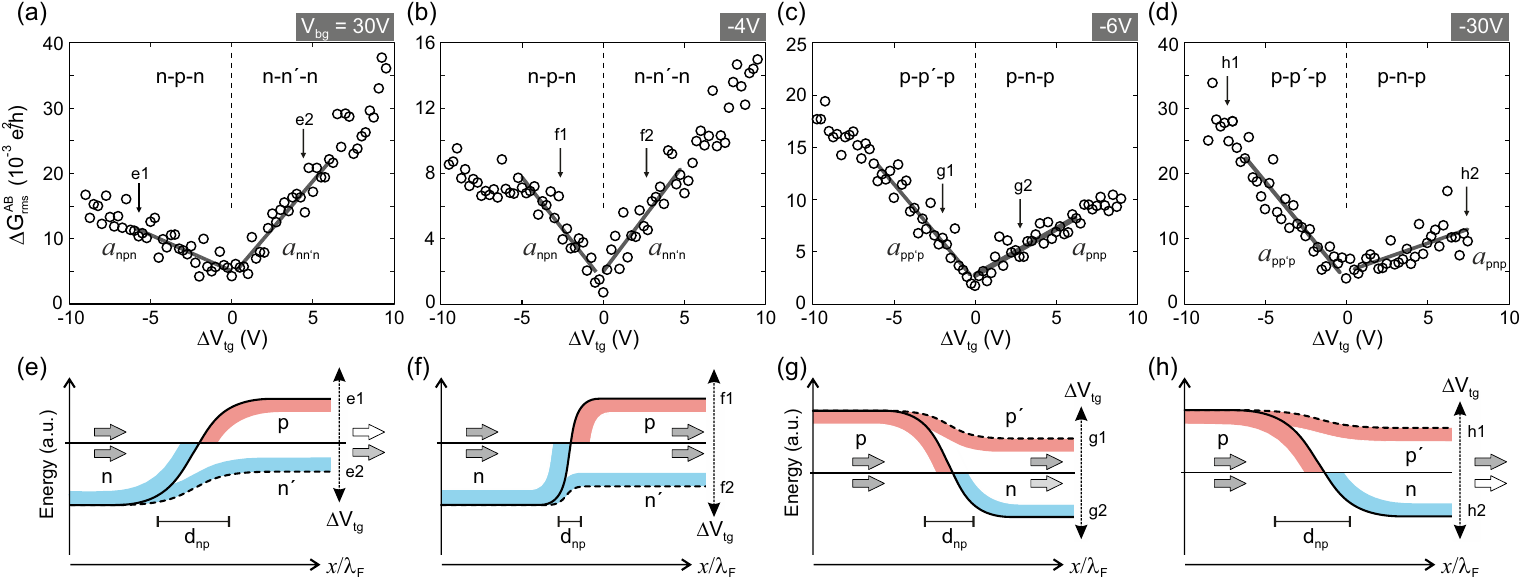}                   
\caption{
Aharonov-Bohm oscillation amplitudes and Klein tunneling. \textbf{(a)}  $\Delta G_{\text{rms}}^{\text{AB}}$ as a function of the top gate voltage $\Delta V_{\text{tg}}$ for  fixed  $V_{\text{bg}}$ (see label) highlighting the cross-over from the npn to the nn$'$n regime. The gray lines are linear fits. Note that each data point is obtained by an analysis akin to Fig.~\ref{figure2}. \textbf{(b)}-\textbf{(d)} Similar data as in panel (a) but for different back gate voltages (see labels) and thus different $E_\text{F}$ allowing to enter the regime of $\lambda_\text{F}>w_{\text{pn}}$ (panel (b)).  Panels \textbf{(e)}-\textbf{(h)} are schematic illustrations outlining the different junction configurations for the panel above [note the corresponding labels e1, e2, \dots, h2 in panels (a)--(d) as well as in panels (e)--(h)].}
\label{figure3}
\end{figure*}

Figure~\ref{figure1}e shows a color plot of the four-terminal conductance $G_{\text{4T}}$ as function of $V_{\text{bg}}$ and $V_{\text{tg}}$ at a base temperature of $T=36$~mK. The regions of suppressed conductance (dark color) are along two straight lines, which correspond to  (i) $E_\text{F} \approx 0$, where the carrier density in the leads and the upper ring arm are tuned near charge neutrality (gray dashed line), or (ii) $E_\text{F} \approx U_0$, where the carrier density in the lower ring arm is tuned near the charge neutrality point (white dashed line).
Thus, the two dashed lines separate regions with and without pn-junctions at the entrance and exit area of the lower ring arm (see labels in Fig.~\ref{figure1}e). When tuning $V_{\text{bg}}$ and $V_{\text{tg}}$ such that the carrier density in the entire device is rather homogeneous, we observe the graphene characteristic piece-wise linear conductance response (Fig.~\ref{figure1}f) 
\rev{as well as the emergence of Landau levels as function of magnetic field (Fig.~\ref{figure1}g).  }
In good agreement with previous experiments on etched hBN/graphene/hBN samples where scattering at the sample edges is the dominant scattering mechanism~\cite{Terres2016,Dauber2017Nov}, we observe the appearance of the quantum Hall effect exactly from the limit where the cyclotron radius $r_\mathrm{C}(B,n)=\hbar \sqrt{\pi n}/(eB)$ is equal to half the ring arm width ($r_\mathrm{C}=w/2$), as highlighted by the dashed line in Fig.~\ref{figure1}g.
This, together with the large (order $e^2/h$) and fully reproducible phase-coherent conductance fluctuations (see inset in Fig.~1f), proves that we are in the regime of quasi-ballistic transport, $w<l_m<\pi \overline{r}$, where $l_m$ is the mean free path.
Please note that this is in great contrast to earlier experiments on locally gated graphene rings~\cite{Huefner2010,Smirnov2012}.

The phase-coherent character of \rev{the conductance} fluctuations becomes even more apparent, when looking at the magnetic field dependency for fixed gate voltages as shown in Fig.~\ref{figure2}a. Besides weak localization (conductance dip at $B~=~0$~T) and universal conductance fluctuations the data also contain periodic oscillations which arise from the AB effect. The latter becomes more visible when subtracting the averaged conduction-background (see red trace in Fig.~\ref{figure2}a and Supp. Material) resulting in $\Delta G$, which clearly shows AB oscillations (Fig.~\ref{figure2}b). The largest period of the AB oscillations is about $\Delta B \approx 5.8$~mT, in good agreement with our ring geometry ($\Delta B=h/(\pi e \overline{r}^2) \approx 5.3$~mT). In Fig.~\ref{figure2}c we show the Fourier spectrum of $\Delta G$ obtained by fast Fourier transformation (FFT). These data exhibit the presence of several characteristic frequencies. While the lowest $\Delta B$ frequency matches well the fundamental mode of the AB oscillations, the observed higher frequencies agree with the geometrically allowed higher-order contributions (see arrows in Fig.~\ref{figure2}c). From the peaks in the FFT spectrum we estimate the phase coherence length $l_{\phi}$ by analyzing the ratio of the AB amplitudes of the different modes, resulting in $l_{\phi} \approx 1.5~\mu$m for the data presented in Fig.~2c (see inset).
Repeating this type of measurements for different \rev{$V_{\text{bg}}$ and $V_{\text{tg}}$},
we observe AB oscillations similar to the ones shown in Fig.~\ref{figure2}b for all tested gate-voltage configurations. 
\rev{In Fig.~3a we show the background subtracted conductance at a constant $V_{\text{bg}}$ but varying $V_{\text{tg}}$ demonstrating the 
gate dependence of the AB oscillation amplitude. 
Note that this is in great contrast to earlier experiments on a diffusive top-gated graphene ring, where no top gate dependence has been observed~\cite{Smirnov2012}}.

\rev{The high quality and quasi-ballistic nature of the studied sample is also reflected in the bias voltage ($V_\text{DC}$) dependence of the root mean square (RMS) of the  AB oscillations, $\Delta G_{\text{rms}}^{\text{AB}}$.
At sufficiently low temperature the energy smearing $eV_\text{DC}$ of the quantum state eigenenergies leads to phase decoherence. For energies exceeding the Thouless energy $E_c$ and in absence of inelastic scattering this results in a decrease of $\Delta G_{\text{rms}}^{\text{AB}} \propto \sqrt{E_c/(eV_\text{DC})}$~\cite{Ren2013}. This is in good agreement with experiments as highlighted by the dashed line in Fig.~3c, where the threshold voltage $E_c/e=\hbar v_\mathrm{F}/(e\,l_m)$ given for ballistic transport ($l_m \sim \pi \overline{r}$) fits reasonably well (see arrow in Fig.~3c).}

\rev{In our theoretical analysis, we assume that the transmission through the device is dominated by the fundamental mode, as $l_\phi$ approximately equals $\pi \overline{r}$ and we assume that the mean free path $l_m$ in the device is larger than the width $w_\mathrm{pn}$ of the pn-junction. The latter implies that we can describe tunneling through the junction by a tunneling amplitude.}

Using this semiclassical framework to describe transport through quasi-ballistic graphene rings, we find that $\Delta G_{\text{rms}}^{\text{AB}}$ is approximately given by (see Supp. Materials)
\begin{equation}
    \Delta G_{\text{rms}}^{\text{AB}} = \frac{4\sqrt{2} e^2}{h}   \; N(E_\text{F}) \; |t_{\text{pr}}|,
\end{equation}
where $N(E_\text{F})$ is the number of modes in the lead
and $|t_{\text{pr}}|$ is the product, averaged over all modes, of the absolute values of the tunneling amplitudes $|t_{\text{up}}|$ and $|t_{\text{lo}}|$ through the upper (up) and lower (lo) ring arm. 
Because of the geometry of the device, $|t_{\text{up}}|$ depends only
\rev{ on number of modes set by the Fermi energy $E_\text{F}$.
On the other hand, $|t_\mathrm{lo}|$ also depends $U_0$, which is tuned by the top gate, as it affects the number of modes in the lower ring arm.} 
By investigating $\Delta G_{\text{rms}}^{\text{AB}}$, we  can thus study $|t_{\text{pr}}|$ as function of the barrier height $U_0$ while keeping all other parameters fixed. As these measurements are repeated for different $E_\text{F}$, we are able to study the transparency of the pn-junction over a wide range of parameters.


\begin{figure}[tb]\centering 
\includegraphics[draft=false,keepaspectratio=true,clip,%
                   width=0.86\linewidth]%
                   {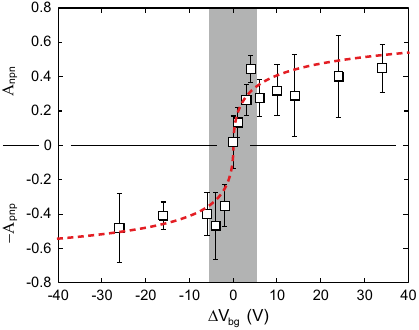}                   
\caption{Tunability of the effective width of the pn-junction.
The normalized difference $A_\text{npn}$ ($-A_\text{pnp}$) of the slopes $a_\text{nn$'$n}$ and $a_\text{npn}$ ($a_\text{pp$'$p}$ and $a_\text{pnp}$) as function of $\Delta V_{\text{bg}} = V_{\text{bg}}- V_{\text{bg}}^0$, where $V_{\text{bg}}^0$ denotes the charge neutrality point in back gate voltage. The shaded area highlights the regime with $d_{\text{pn}}<1$, corresponding to $\lambda_\text{F}>w_{\text{pn}}$. The dashed line results from the parameterless semiclassical model [Eq.~\eqref{eq:Anpn}].
} 
\label{figure4}
\end{figure}

To investigate the transmission through the ring as function of $U_0$ and $E_\text{F}$, we  plot the amplitude of the AB oscillations $\Delta G_{\text{rms}}^{\text{AB}}$ as a function of 
$\Delta V_{\text{tg}} = V_{\text{tg}} - V^0_{\text{tg}}$ for several values of $V_{\text{bg}}$,
as shown in Figs.~\ref{figure3}a-d. Here, $V^0_{\text{tg}}$ denotes the charge neutrality point \rev{(CNP)} at fixed $V_\text{bg}$. Changing   $V_{\text{bg}}$ modifies $\lambda_\text{F}$  and therefore the effective width $d_\text{pn}$ of the  pn-junctions. 
As 
 observed in  Figs.~\ref{figure3}a-d, this has a large impact on the slope of $\Delta G_{\text{rms}}^{\text{AB}}$ as function of $\Delta V_{\text{tg}}$. 
In Figs.~3a-b, $\Delta G_{\text{rms}}^{\text{AB}}$ is largest for the largest positive values of $\Delta V_{\text{tg}}$, when the lower arm is in the nn$'$n-regime. The amplitude  decreases when $\Delta V_{\text{tg}}$ is reduced, it reaches a minimum for  $\Delta V_{\text{tg}}=0$ ($E_\text{F} \approx U_0$), and it increases again for $\Delta V_{\text{tg}}<0$, when the lower arm forms a  npn-junction. This behavior is due to the fact that reducing $\left| \Delta V_{\text{tg}} \right|$ decreases the number of open modes in the lower ring-arm, lowering $|t_{\text{pr}}|$. What is remarkable is that, while in Fig.~\ref{figure3}a the slope of the data points is much smaller for $\Delta V_{\text{tg}}<0$ than for $\Delta V_{\text{tg}}>0$, the situation is fairly symmetric in Fig.~\ref{figure3}b \rev{for $|\Delta V_{\text{tg}}|<5$~V.
Note that in the regime $\Delta V_{\text{tg}}<-5$~V we are limited due to a conductance drop in the lead regions. This is due to the (weak but finite) capacitive coupling of the top gate with the lead regions (see slight negative slope of the gray dashed line in Fig.~1e) and becomes important for $V_{\text{bg}}$ values close to the CNP (see also Supp.~Fig.~S11).
Moreover, we observe that $\Delta G_{\text{rms}}^{\text{AB}}$ scales also with $V_{\text{bg}}$, i.e. the carrier density which is in agreement with eq.~(1). 
} 

The difference between the configurations of Fig.~\ref{figure3}a and Fig.~\ref{figure3}b is that in the second case $V_{\text{bg}}$ is close to the \rev{CNP}, implying that the effective width of the junctions is small, $d_{\text{pn}} \lesssim 1$. 
Such sharp junctions are almost transparent for the charge carriers (Klein tunneling), so that the behavior is symmetric (for small $|\Delta V_{\text{tg}}|$ values) in the nn$'$n and npn sides. 

\rev{
Vice versa, in the configuration of Fig.~\ref{figure3}a, $V_\text{bg}$ is away from the CNP, so that the effective width of the junctions is large, $d_\mathrm{pn} > 1$. For $\Delta V_\mathrm{tg} < 0$, the overall transmission probability through the lower ring-arm is reduced due to collimation at the two pn-interfaces.
}
The situation is similar for what concerns the amplitude of the AB oscillation when the lower ring-arm is in the pp$'$p or in the pnp regime, see Figs.~\ref{figure3}c-d. 
Also in this case the asymmetry of $\Delta G_{\text{rms}}^{\text{AB}}$ around
$\Delta V_{\text{tg}}=0$ increases for $V_{\text{bg}}$ values further away from the \rev{CNP} because of the reduced transmission of wider pn-junctions.    
For example, when comparing the values at the same absolute carrier density (see e.g. labels h1 and h2 in Fig.~\ref{figure3}d) we extract a 35\% reduced transparency for the pnp-junctions.

In order to better understand this behavior, we use a semiclassical model to calculate the tunneling amplitude $|t_\text{pr}|$ (see Supp. Materials).
Approximating the potential of the pn-junctions by a linear potential~\cite{Cheianov2006,Tudorovskiy12} (see dotted line in Fig.~1d), we obtain (see Supp. Materials) 
\begin{equation}
  |t_{\text{pr}}| = \sqrt{\frac{U_0}{8 \pi d_{\text{pn}}E_\text{F}}}\;
   \;\operatorname{Erf}\left( \sqrt{\frac{2\pi^2 d_{\text{pn}}}{U_0 E_\text{F}}} \left(U_0-E_\text{F}\right) \right),
\end{equation}
valid for $U_0 \leq 2 E_\text{F}$, where $\operatorname{Erf}(x) = (2/\sqrt\pi)\int_{0}^x e^{-t^2}dt$ denotes the error function. We observe that increasing $U_0$ not only increases the number of open modes in the lower ring arm, but also the transmission probability through the available open modes (second and first factor in Eq.~(2), respectively). As a combination of these two mechanisms, the npn-junction becomes highly transparent for $V_{\text{bg}}$ near \rev{the CNP}. 
Using the Eqs. (1) and (2), we can calculate the slope of $\Delta G_{\text{rms}}^{\text{AB}}$ as function of  $\Delta V_{\rm tg}$ in the various regimes npn, nn$'$n, pnp and pp$'$p for each value of $E_\text{F}$.  Focusing on the normalized difference of the slopes $a_\text{nn$'$n}$ and $a_\text{npn}$, we  obtain
\begin{equation}\label{eq:Anpn}
  A_\text{npn} = \frac{|a_\text{nn$'$n}|-|a_\text{npn}|}{|a_\text{nn$'$n}|+|a_\text{npn}|} =  \frac{1-f(d_{\text{pn}})}{1+f(d_{\text{pn}})},
\end{equation}
where $ f(d_{\text{pn}}) = \sqrt{3 /(4\pi\;d_{\text{pn}})} \; \operatorname{Erf}\bigl(\pi \sqrt{ d_{\text{pn}} / 3 } \bigr)$ (see Supp. Material). For the pnp-regime, we need to exchange the lables $\text{n}\leftrightarrow\text{p}$. Importantly, this result depends only on $d_{\text{pn}}=k_\text{F} w_{\text{pn}}/2\pi$ and does not contain any \rev{other} parameter.
\rev{By taking the geometrically fixed width of the pn-junction as $w_{\text{pn}} = 60$~nm 
(see above) we find excellent agreement between theory and experiment (see Fig.~\ref{figure4}), where the slopes have been extracted from the data shown in Fig.~\ref{figure3} and Supp. Fig.~S11.}
%
This indicates that: (a) our model captures the essential physics of the system, (b) that the regime $\lambda_\text{F}>w_{\text{pn}}$ can be well achieved in our device, and, (c) that we can tune the pn-junction from efficient Klein-tunneling (high transmission) to low transmission in a controlled manner.

This result is an explicit confirmation of the long-predicted tunability of pn-junctions in graphene devices and shows that, by combining clever device-design and high-quality graphene/hBN heterostructures, it is possible to realize mesoscopic devices that go well beyond what can be achieved with conventional two-dimensional electron gases.
For example, this together with highly transparent superconducting contacts~\cite{Calado2015Jul} and soft-etching (as demonstrated for ballistic anti-dot lattices~\cite{Jessen2019Feb}) opens the door to highly tunable Andreev billiards and many other exotic coherent electron optics devices.

\section*{Aknowledgements}
We thank J.~G\"uttinger, H.~Bluhm and F.~Haupt for helpful discussions.
This project has received funding from the European Union's Horizon 2020 research and innovation programme under grant agreement No 881603 (Graphene Flagship) and from the European Research Council (ERC) under grant agreement No. 820254, the Deutsche Forschungsgemeinschaft (DFG, German Research Foundation) under STA 1146/11-1, and by the Helmholtz Nano Facility~\cite{Albrecht2017May}.  
K.W. and T.T. acknowledge support from the Elemental Strategy Initiative conducted by the MEXT, Japan, Grant Number JPMXP0112101001, JSPS KAKENHI Grant Number JP20H00354 and the CREST(JPMJCR15F3), JST.
K.J.A.R. and M.I.K. acknowledge support from the Netherlands Organisation for Scientific Research (NWO) via the Spinoza Prize.

 Supplementary Information accompanies the online version of this paper.

\bibliography{diss}

\end{document}



\baselineskip24pt


\maketitle 

\newpage

%

%
%


\section{Supplementary details on experiment}

\subsection*{Sample fabrication}
The heterostructure was made from CVD grown graphene and mechanically exfoliated hBN by van der Waals assembly 
and was placed on a highly doped Si substrate with a $285$~nm dry thermally grown SiO$_2$ top layer. The structure was patterned and contacted using standard electron beam lithography, reactive ion etching with a SF$_6$/Ar plasma and electron beam evaporation. An Al hard mask was used for etching. Contacts were made of Cr/Au ($5$\,nm/$95$\,nm).  The structured and contacted device was covered with an additional hBN flake as gate dielectric and subsequently a partial top gate was built with similar processes as for the contacts,  but with Cr/Au ($5$\,nm/$145$\,nm). Before measurements the device was heat annealed in Ar/H$_2$ atmosphere at $300^\circ$C for several hours.

\subsection*{Device characterization}

\begin{figure}[h]\centering
\includegraphics[draft=false,keepaspectratio=true,clip,width=1\linewidth]{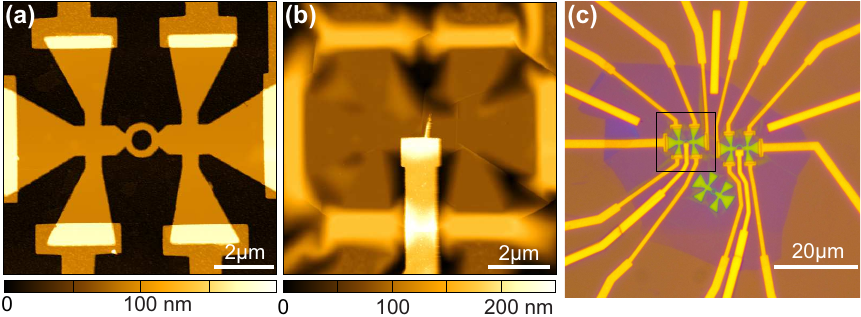}
\caption[figS1]{ (a) SFM image of patterned and contacted device without top-gate hBN. (b) SFM image of the device after additional hBN flake transfer and top-gate metallization. (c) Optical picture of the final device studied in this work (marked by black box).}
\label{figS1}
\end{figure}

Fig.~\ref{figS1} displays optical and scanning force microscopy (SFM) images of different stages in the device fabrication. The patterned and contacted hBN/graphene/hBN heterostructure is shown in Fig.~\ref{figS1}(a). From this SFM image the dimensions of the ring were extracted ($\overline{r} = 500$\,nm, $w = 200$\,nm).  Fig.~\ref{figS1}(b) depicts the device after the transfer of the top gate hBN flake and top gate metallization. The top gate hBN flake adapted to the structure only to some extent and is partially suspended between metal contacts and etched heterostructure. The top gate followed the uneven surface, but the metal thickness was chosen sufficiently large to overcome the appearing height differences (top-gate metal Cr 5\,nm/Au 145\,nm ). At the top of the metal finger a spike appears, most likely an artifact of the lift-off process in the frame of top-gate metallization, but it had no effect on the top-gate functionality. An optical image of the final device is displayed in Fig. \ref{figS1}(c). All measurements were performed in wet dilution refrigerator with perpendicular magnetic field and at a base temperature of $36$~mK, unless stated otherwise. We used low-frequency AC lock-in techniques for simultaneous two and four-terminal measurements with a constant AC bias of $V_{\text{AC}}=200\,\mu$V at $83.22$\,Hz. Bias dependency is investigated by different bias voltages $V_{\text{b}}=V_{\text{AC}}+V_{\text{DC}}$, where a constant AC bias $V_{\text{AC}}=100~\mu$V is overlayed with a variable DC bias $V_{\text{DC}}$.  

\begin{figure}[h]\centering
\includegraphics[draft=false,keepaspectratio=true,clip,width=1\linewidth]{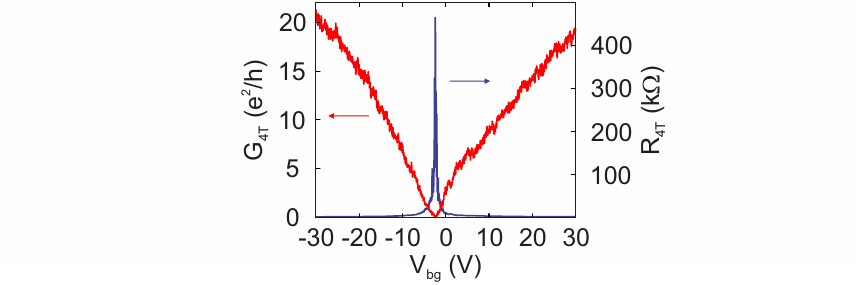}
\caption[figS2]{$G_{\text{4T}}$ and $R_{\text{4T}}$ versus $V_{\text{bg}}$ for fixed relation to top-gate voltage $V_{\text{tg}}=(V_{\text{bg}}+2.5 V)/4$ providing similar charge carrier densities in both ring arms.}
\label{figS2-1}
\end{figure}

Fig. \ref{figS2-1} shows the four-terminal conductance $G_{\text{4T}}$ and resistance $R_{\text{4T}}$ as function of back-gate voltage $V_{\text{bg}}$ with a fixed relationship to top-gate voltage $V_{\text{tg}}=(V_{\text{bg}}-V_{\text{bg}}^{0})/\beta$. The position of the charge neutrality point (CNP), $V_{\text{bg}}^{0}=-2.5$\,V, and the slope $\beta=4$ are extracted by adjusting both parameters until the most linear $G_{\text{4T}}$ and the sharpest peak of $R_{\text{4T}}$ are found. Thus, the contributions from the two ring arms are aligned at best and only a single Dirac peak is visible in the gate characteristics. The slope $\beta=4$ is in good agreement with the distance ratio between the back-gate to the graphene ($\approx 285$\,nm) and the top-gate to the graphene ($\approx 60$\,nm). From the color plot of $G_{\text{4T}}$ as function of $V_{\text{bg}}$ and $V_{\text{tg}}$  (see main text Fig. 1e) we extract a lever arm ratio of $\alpha_{\text{tg}}/\alpha_{\text{bg}}\approx 4.4$, which is in good accordance with the determined slope. With this relationship nearly equal charge carrier densities are achieved in both ring arms.

The same relationship of $V_{\text{bg}}$ and $V_{\text{tg}}$ is used to study the magneto-conductance of the device by quantum Hall measurements. We observe a graphene-typical Landau fan with integer Hall plateaus and an absent zero Landau level. At small, positive $V_{\text{bg}}$ a bending of the Landau levels occurs, which coincides with a non-linearity in the gate characteristics (compare Fig.~\ref{figS2-1}). This effect could arise from an inhomogeneous doping profile or localized state at the edge of the etched structure in combination with the complex electrostatic tuning of the device, but it is not fully understood. Following Ref. \cite{Ki2012} we extract a back-gate lever arm $\alpha_{\text{bg}}\approx 6.4\times 10^{10}$~cm$^{-2}$V$^{-1}$ over an average of filling factors $\nu=-10,-6$ and $-2$, which is in good agreement with a parallel plate capacitor model ($\alpha \approx 7.1\times 10^{10}$~cm$^{-2}$V$^{-1}$, $\epsilon_r^{\text{hBN}}=4$).

\begin{figure}[h]\centering
\includegraphics[draft=false,keepaspectratio=true,clip,width=1\linewidth]{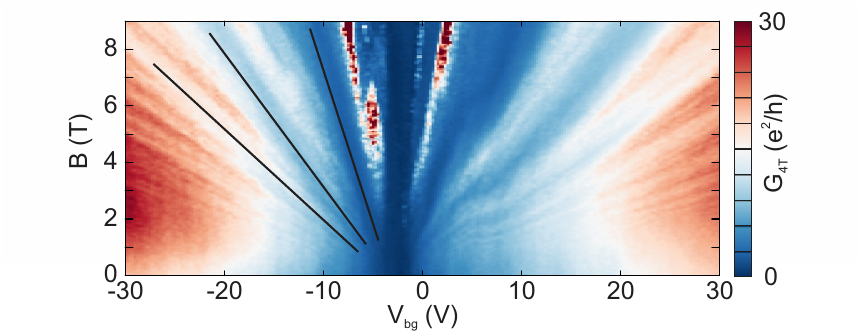}
\caption[figS2]{Four-terminal conductance as function of back-gate voltage $V_{\text{bg}}$ and magnetic field $B$. The top-gate voltage is swept following $V_{\text{tg}}=(V_{\text{bg}}+2.5V)/4$. Black solid lines mark slopes for the extraction of the lever arm $\alpha_{\text{bg}}$.}
\label{figS2}.
\end{figure}


\begin{figure}[h]\centering
\includegraphics[draft=false,keepaspectratio=true,clip,width=0.5\linewidth]{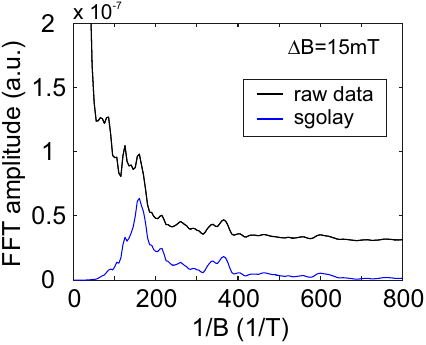}
\caption[figS8]{ Fourier transform of four-terminal magneto-conductance before (black) and after (blue) background subtraction. Curves are offset. }
\label{figS8}
\end{figure} 

\newpage
\subsection*{Background subtraction and Fourier transform} 

Data post-processing is crucial to separate AB oscillations from other transport effects in mesoscopic systems, e.g. weak localization, universal conductance fluctuations (UCFs) and size effects. The magneto-conductance background beside UCFs changes rather slowly compared with the periodicity of the AB oscillations. Therefore these effects are easily filtered out by averaging. 

The periodicity of AB oscillations is given by sample geometry $\Delta B_j(r)= h/(m\,e\,\pi\,r^2$) with $m=1,2,...$, where $m$ is the mode number. With our ring geometry $\overline{r}=500$~nm and $w=200$~nm we calculate for the fundamental mode $\Delta B_1(\overline{r}-w/2)=3.7$\,mT, $\Delta B_1(\overline{r})=5.3$\,mT and $\Delta B_1(\overline{r}+w/2)=8.2$\,mT and the first harmonic $\Delta B_2(\overline{r}-w/2)=1.8$\,mT, $\Delta B_2(\overline{r})=2.6$\,mT and $\Delta B_2(\overline{r}+w/2)=4.1$\,mT. 
All other higher harmonics are calculated in a similar manner. This periodicity in magnetic field translates into an expected frequency range of $(121-273)\,1/$T for $h/e$, $(243-550)\,1/$T for $h/(2e)$ and $(364-913)\,1/$T for $h/(3e)$ in the Fourier spectrum. 
The periodicity of UCFs in a quasi-diffusive system can be estimated by $\Delta B_{\text{UCF}} \approx h/(e\,w\, l_\phi)$~\cite{Huefner2010}, where $w$ is the  width of the ring and $l_\phi$ the phase-coherence length. 
Taking $l_\phi=1.5$\,$\mu$m (see main text and below) this estimate gives $\Delta B_{\text{UCF}}=13.8$\,mT or a frequency of $72\,$T$^{-1}$, respectively. This frequency is clearly distinguishable from the expected AB oscillations and can be separated with an adequate filter method. 

The magneto-conductance background $\left\langle G_{\text{4T}} \right\rangle_{\Delta B}$ is determined by filtering the four-terminal conductance $G_{\text{4T}}$ over the span $\Delta B$. The background subtracted conductance $\Delta G = G_{\text{4T}} - \left\langle G_{\text{4T}} \right\rangle_{\Delta B}$ is calculated by subtracting the filtered data from the raw data and is used for further analysis of the AB oscillations. We choose a Savitzky-Golay filter~\cite{Savitzky1964Jul} with a 5th order polynomial and a span of approximately three periods of $h/e$ oscillations ($15$~mT) for the extraction of the magneto-conductance background $\left\langle G_{\text{4T}} \right\rangle_{\Delta B}$. 
This method provides higher filter dynamic and steeper filter edge compared to a moving average without distorting the signal tendency. In Fig.~\ref{figS8} we compare the Fourier transform of the raw data with the Fourier transform of the background subtracted conductance $\Delta G$ determined by the Savitky-Golay filter. 
Importantly, even in the raw data the frequency components of the AB oscillations and the first harmonic are recognizable, but overlapped with an exponentially decaying frequency background. 
The Fourier transform of the processed signal reproduces the very same features of AB oscillations as shown in the raw data and in addition removes the frequency background.


\begin{figure}[h!]\centering
\includegraphics[draft=false,keepaspectratio=true,clip,width=0.8\linewidth]{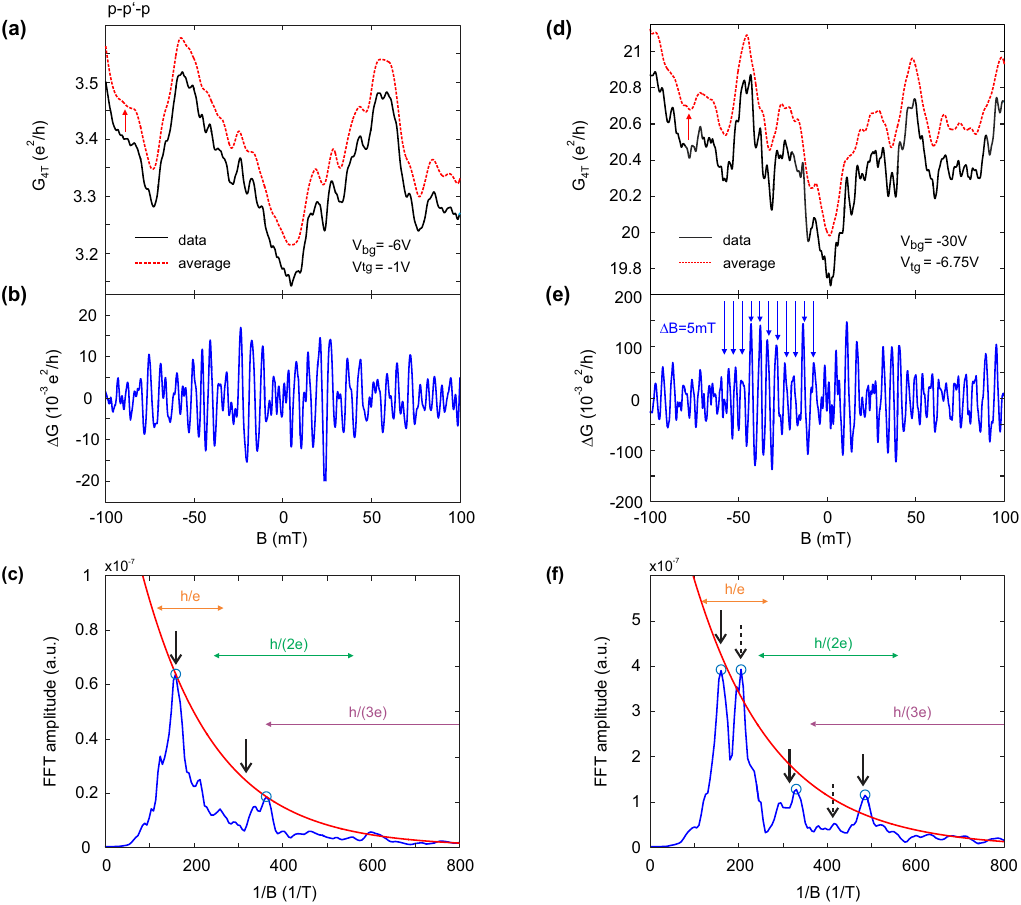}
\caption[figS12]{AB oscillations for different gate configurations: (a)-(c) p-p'-p regime close to CNP and (d)-(f) high gate voltages in p-p'-p regime. (a) and (d) $G_{\text{4T}}$ as function of $B$-field. Dashed lines represent the smoothed data used for background subtraction (plotted offset). (b) and (e) $\Delta G$ extracted from the data presented in panel (a) and (d), respectively. 
(c) and (f) Corresponding Fourier transform with indicated frequency ranges of AB modes. Arrows mark observed fundamental modes and their expected higher harmonics. Red solid line represents a exponential fit to maxima of the peaks (blue circles) for the estimation of $l_\phi$.}
\label{figS12a}.
\end{figure}

\begin{figure}[h!]\centering
\includegraphics[draft=false,keepaspectratio=true,clip,width=0.95\linewidth]{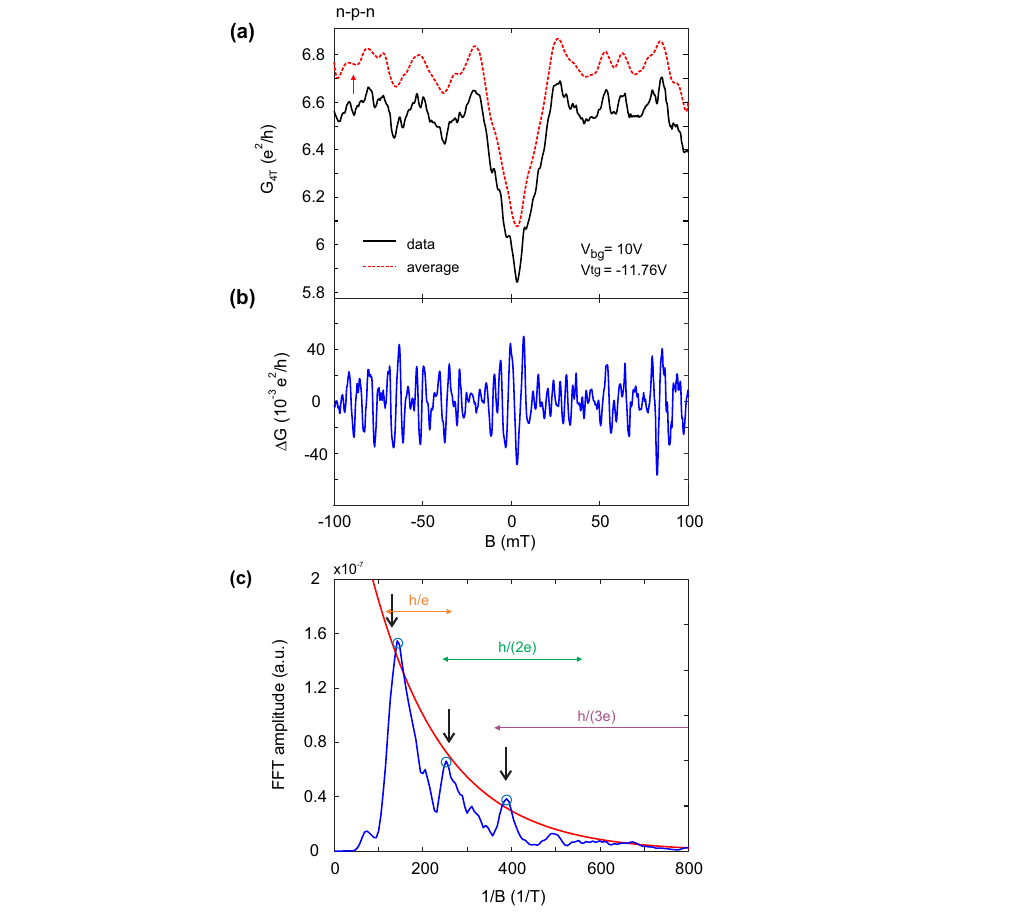}
\caption[figS12a]{Similar plot as in Fig.~\ref{figS12a} but for a n-p-n gate configurations:(a) $G_{\text{4T}}$ as function of $B$-field. Dashed lines represent the smoothed data used for background subtraction (plotted offset). (b) $\Delta G$ extracted from the data presented in panel (a) and (d), respectively. 
(c) Corresponding Fourier transform with indicated frequency ranges of AB modes. Arrows mark observed fundamental modes and their expected higher harmonics. Red solid line represents a exponential fit to maxima of the peaks (blue circles) for the estimation of $l_\phi$.}
\label{figS12ab}.
\end{figure}

\subsection*{Additional data on Aharonov-Bohm oscillations} 

In the low magnetic field regime, we observe AB oscillations at all gate voltage settings independent of being in an unipolar or bipolar doping regime. In Figs.~\ref{figS12a} and~\ref{figS12ab} we present additional data sets for the observation of AB conductance oscillations. Similar to Fig. 2 in the main text, the raw data of the four-terminal conductance $G_{\text{4T}}$, the background subtracted conductance $\Delta G$ and the Fourier transform of $\Delta G$ are displayed for two different gate configurations. Close to the CNP in the p-p'-p regime (see Fig.~\ref{figS12a}(a-c)) we find AB oscillations of the fundamental mode and the first harmonic and extract a $l_\phi \approx 1.5\,\mu$m. For high charge carrier density in the p-p'-p regime (Fig.~\ref{figS12a}(d-f)) we observe an Fourier spectrum with higher harmonics up to $h/(3e)$ and extract a $l_\phi \approx 1.5\,\mu$m. In general we notice that for higher back-gate voltages and for similar charge carrier densities (and types) in both ring arms the higher harmonics are best visible in the Fourier spectra. With a homogeneous carrier density inside the ring, no pn-junctions or interface are formed, where backscattering could occur. Also, 
with increasing $V_{\text{bg}}$ the mean free path increases making scattering more unlikely. 
Therefore, this regime is ideal for the observation of AB oscillations and their
higher harmonics.
Fig.~\ref{figS12ab} shows very similar data as shown in Fig. 2 of the main manuscript and in Fig.~\ref{figS12a} but in the n-p-n regime.



\begin{figure}[h!]\centering
\includegraphics[draft=false,keepaspectratio=true,clip,width=1\linewidth]{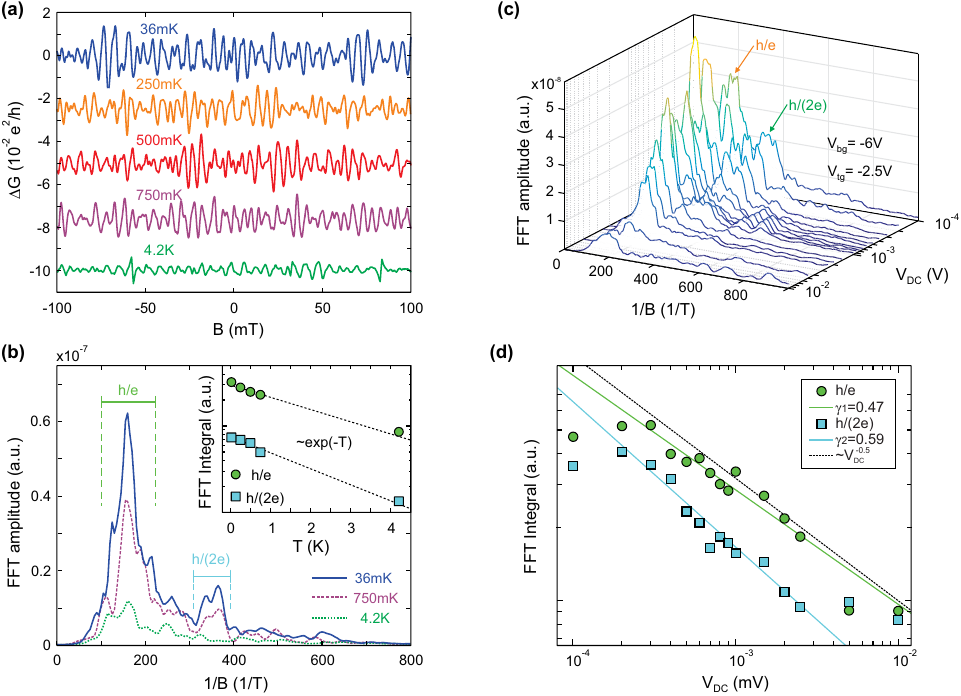}
\caption[figS10]{Temperature and bias voltage dependency of AB oscillations. (a)  background subtracted conductance as a function of magnetic field for various temperatures $T$ at $V_{\text{bg}}=-6$~V and $V_{\text{tg}}=-1$~V with constant offset for clarity. (b) Fourier transforms of selected traces from panel (a). Insert
shows the integral for $h/e$ and $h/(2e)$ modes over frequency ranges as marked in the plot. Dashed lines represent exponential fits to the data. (c) Fourier transform of $\Delta G$ as a function of magnetic field and DC bias voltage at $V_{\text{bg}}=-6$~V and $V_{\text{tg}}=-2.5$~V. (d) AB amplitude plotted as a function of DC bias voltage of data shown in panel (c). The solid lines represent a fit $\propto V^{-\gamma_m}$ and the black dashed line is a guide to the eye ($\propto 1/\sqrt{V_{\text{DC}}}$).} 
\label{figS08a}
\end{figure} 

\subsection*{Temperature and bias dependent measurements}

Fig.~\ref{figS08a}(a) shows AB oscillations measured at different temperatures $T$. The amplitude decreases with temperature, while the periodicity is preserved. The Fourier transform of this data is displayed in Fig.~\ref{figS08a}(b), where the $h/(2e)$ component is visible up to $T=750$~mK and the $h/e$ component is present even at $T=4.2$~K. Besides the decrease of amplitude slightly-different peak structures are observed with changing temperature. 
For a quantitative comparison, we calculate the integral of the $h/e$ and $h/(2e)$ peaks over ranges as indicated in Fig.~\ref{figS08a}(b) (see vertical dashed lines). The data is well described by an exponential decay (see insert of Fig.~\ref{figS08a}(b)). 
Decoherence of AB oscillations is studied further by DC bias voltage $V_{\text{DC}}$ dependent measurements, presented in a waterfall plot of the Fourier transform of $\Delta G$ as a function of inverse magnetic field ($1/B$), as shown in Fig.~\ref{figS08a}(c).  
The amplitudes of the $h/e$ and the $h/(2e)$ mode decrease with increasing $V_{\text{DC}}$ and are nearly vanished at $V_{\text{DC}}=10$~mV and $2$~mV, respectively. This decrease is quantified in Fig.~\ref{figS08a}(d), where 
the integral of the FFT amplitude over the two regimes (see vertical dashed lines in Fig.~\ref{figS08a}(b)) is plotted as function of $V_{\text{DC}}$ in a double logarithmic plot exhibiting a $1/\sqrt{V_{\text{DC}}}$ dependency.
Please note that a very similar behaviour is also found when plotting directly 
$\Delta G_{\text{rms}}^{\text{AB}}$ as function of $V_{\text{DC}}$ (see Fig.~S10(a)). The corresponding data of $\Delta G$ versus $B$-field for different $V_{\text{DC}}$ are shown in Fig.~S10(b).
%

\begin{figure}[h]\centering
\includegraphics[draft=false,keepaspectratio=true,clip,width=0.88\linewidth]{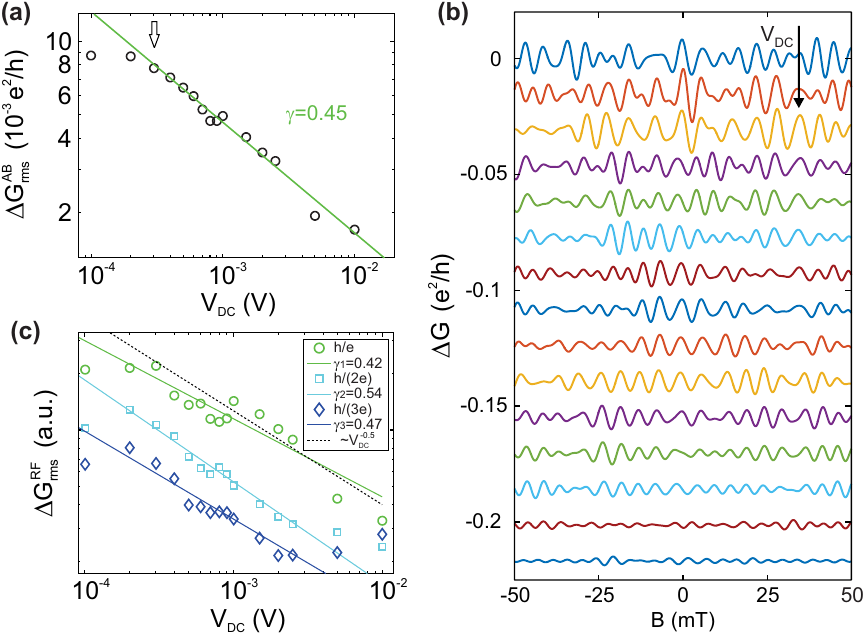}
\caption[figS3]{Additional analysis of DC bias dependency of AB amplitude at $V_{\text{bg}}=-6\,$V and $V_{\text{tg}}=-2.5\,$V: (a) RMS of background subtracted conductance, $\Delta G_{\text{rms}}^{\text{AB}}$ as function of DC bias.
(b) corresponding $\Delta G$ as function of $B$-fields for different $V_{\text{DC}}$ (for range see $x$-axis in panel (a)).(c) RMS of the conductance from the reverse Fourier transform for different modes $\Delta G_{\text{rms}}^{\text{RF,m}}$. Solid lines represent power law fits $\propto V_{\text{DC}}^{-\gamma_m}$ to the individual data sets. Black dashed line indicate a guide to the eye for a $V_{\text{DC}}^{-1/2}$ power law.}
\label{figS3}
\end{figure} 

\begin{figure}[h]\centering
\includegraphics[draft=false,keepaspectratio=true,clip,width=0.8\linewidth]{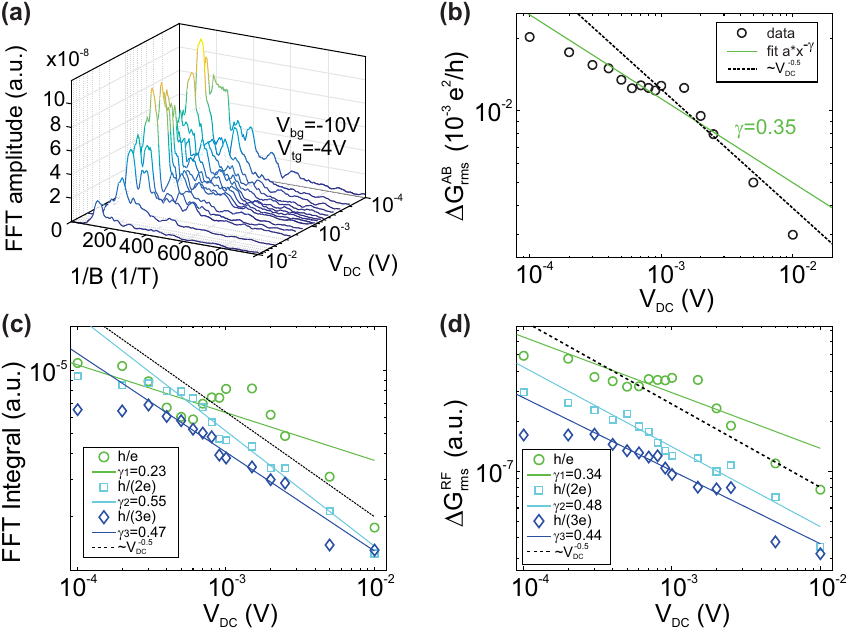}
\caption[figS4]{Additional data on DC bias dependency at $V_{\text{bg}}=-10\,$V and $V_{\text{tg}}=-4\,$V: (a) Fourier transform as function of DC bias voltage. (b) RMS of background subtracted conductance, $\Delta G_{\text{rms}}^{\text{AB}}$ as function of DC bias. (c) Integral of Fourier spectrum over the ranges of AB $h/e$, $h/(2e)$ and $h/(3e)$ modes. (d) RMS of the conductance from the reverse Fourier transform for different modes, $\Delta G_{\text{rms}}^{\text{RF,m}}$. Solid lines represent power law fits $\propto V_{\text{DC}}^{-\gamma_m}$ to the individual data sets. Dashed lines indicate a guide to the eye for a $V_{\text{DC}}^{-1/2}$ power law dependence.}
\label{figS4}
\end{figure} 

\begin{figure}[h]\centering
\includegraphics[draft=false,keepaspectratio=true,clip,width=0.8\linewidth]{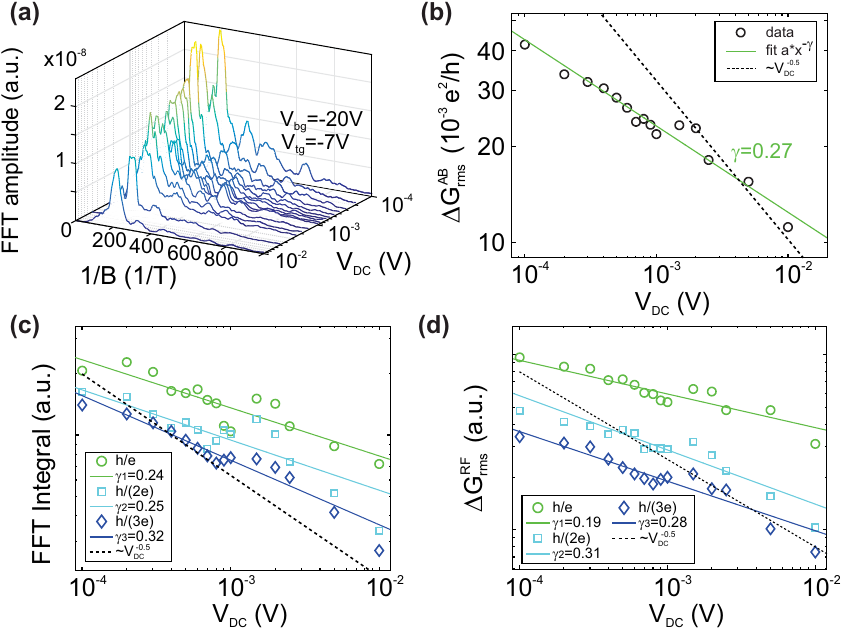}
\caption[figS5]{Additional data on DC bias dependency at $V_{\text{bg}}=-20\,$V and $V_{\text{tg}}=-7\,$V: (a) Fourier transform as function of DC bias voltage. (b) RMS of background subtracted conductance, $\Delta G_{\text{rms}}^{\text{AB}}$ as function of DC bias. (c) Integral of Fourier spectrum over the ranges of AB $h/e$, $h/(2e)$ and $h/(3e)$ modes. (d) RMS of the conductance from the reverse Fourier transform for different modes, $\Delta G_{\text{rms}}^{\text{RF,m}}$. Solid lines represent power law fits $\propto V_{\text{DC}}^{-\gamma_m}$ to the individual data sets. Dashed lines indicate a guide to the eye for a $V_{\text{DC}}^{-1/2}$ power law dependence.}
\label{figS5}
\end{figure}

Two mechanisms lead to phase decoherence at sufficient low temperatures: (i) Energy smearing $\Delta E$ of the quantum state eigenenergies and (ii) changes of the phase coherence length $l_\phi$ affecting the amplitude of AB oscillations: 
\begin{equation}
\Delta G_{\text{rms}}^{\text{AB}} \propto \sqrt{(E_c/\Delta E)} \exp{(-l/l_\phi(T))},
\label{eq:ABamp}
\end{equation}
 where $E_c$ is the Thouless energy given by $E_c=\hbar/\tau$ with $\tau$ the typical traversal time~\cite{Ren2013,Russo2008}. Finite temperature $T$ or applied DC bias voltage $V_{\text{DC}}$ lead to an energy smearing of $\Delta E=k_{\text{B}} T = e V_{\text{DC}}$. If $\Delta E$ is larger than $E_c$, $\Delta E/E_c$ uncorrelated energy levels contribute to transport averaging out electron interference effects proportionally to $\sqrt{(E_c/\Delta E)}$~\cite{Ren2013}. We estimate $E_c$, assuming a diffusive \cite{Russo2008} or ballistic system~\cite{Mur2008} with $l=\pi \overline{r}$, of $56~\mathrm{\mu}$eV (diffusive) and $480~\mathrm{\mu}$eV (ballistic), respectively. 
%
%
From the bias dependent measurements we extract a critical bias voltage of $V_c \approx 300~\mu$V (see arrow in Fig.~S10(a)). This value is between the two cases and we conclude that our device is a quasi-ballistic mesoscopic system, where the electron mean free path $l_m$ is between the sample dimensions ($w<l_m<l$).

At low temperatures $l_\phi$ is mainly influenced by electron-electron interaction \cite{Hansen2001} and in two dimensional systems the phase-breaking time $\tau_\phi$ is found to be $\tau_\phi \propto T$ \cite{Altshuler1982}. Below $E_c$ we observe an exponential decay $\propto \exp(-T)$, which again points towards a ballistic ($l_\phi \propto 1/T$) rather than a diffusive ($l_\phi \propto 1/\sqrt{T}$) system similar to experiments in GaAs/(Al,Ga)As heterostructures~\cite{Hansen2001}. Please note that a critical bias voltage of $V_c \approx 300~\mu$V corresponds to a critical temperature of $T_c \approx 3.5$~K and that most of our measurements have been performed well below~$T_c$. 


\subsubsection*{More details and additional data}

For the analysis of the amplitude of the AB oscillations as function of DC bias voltage we used in total three different methods. First we simply take the RMS value of the background subtracted conductance $\Delta G$ given by $\Delta G_{\text{rms}}^{\text{AB}}$, as it was used by many other groups before~\cite{Russo2008,Ren2013}. The second approach (FFT integral) is based on the integration of the Fourier spectrum of $\Delta G$ with respect to different frequency ranges of the various AB modes. Thus, the amplitudes of the $h/e$, $h/2e$ and $h/3e$ mode are extracted individually and only contributions in the specific frequency ranges are taken into account. For the third method (reverse Fourier (RF) RMS) only the frequency components of the $h/e$, $h/(2e)$ and $h/(3e)$ mode are taken from the Fourier transform of $\Delta G$ and individually re-transformed by reverse Fourier transformation. From the reverse Fourier transformed conductance the RMS value, $\Delta G_{\text{rms}}^{\text{RF},m}$, is determined as a measure of the amplitude of the different modes $m$. This procedure combines the simplicity of RMS value with frequency selectivity of the Fourier transformation. Although the three methods are based on quite different means, the obtained results are found to be very similar. The independence of the observed behavior from the specific method proves the functionality of the (simple) amplitude analysis. Since the results are very similar, we choose the FFT integral of the AB mode ranges in the Fourier spectrum as used measure for further AB amplitude and decoherence investigations in temperature and bias dependent measurements. Nevertheless, we show all three different methods for measuring decoherence as function of DC bias voltage and additional data for higher charge carrier densities in Figs. \ref{figS3} to \ref{figS5}. 
While the FFT integral analysis is shown in Fig.~S9(d), Fig.~S10(a) depicts $\Delta G_{\text{rms}}^{\text{AB}}$ and Fig.~S10(c) $\Delta G_{\text{rms}}^{\text{RF},m}$ as function of DC bias voltage at $V_{\text{bg}}=-6$~V and $V_{\text{tg}}=-2.5$~V. Each data set is fitted with a power law $\propto V_{\text{DC}}^{-\gamma_m}$, where $m=1,2,3$ the first, second and third mode of AB oscillations. For $\Delta G_{\text{rms}}^{\text{AB}}$ only a single fit is made (see Fig.~S10(a)). These results validate our findings above and also show a critical DC bias voltage of around $V_c \approx 300\,\mu$V with a $V_{\text{DC}}^{-1/2}$ decay of the AB amplitude afterwards. 
More data of very similar analysis are shown in  Figs.~\ref{figS4} and \ref{figS5}.
Figs.~\ref{figS4}(a) and \ref{figS5}(b) show the waterfall plots of Fourier transforms as function of DC bias voltage and the three amplitude analyses at $V_{\text{bg}}=-10$~V and $V_{\text{tg}}=-4$~V and at $V_{\text{bg}}=-20$~V and $V_{\text{tg}}=-7$~V, respectively. Here, we observe a shift of $V_c$ to higher bias voltages as the charge carrier density increases. Also a decrease of the AB amplitude occurs below $V_c$, but much weaker than $V_{\text{DC}}^{-1/2}$. The shift of $V_c$ could arise from 
 an enhanced mean free path $l_m$ with increasing $V_{\text{bg}}$. 
Hence, the diffusive constant $D=v_{\text{F}}\,l_m/2$ increases and therefore also $E_c$ and $V_c$. The decoherence mechanism below $V_c$ is not fully understood. Without energy smearing of quantum state eigenenergy, changes in $l_\phi$ can influence the AB amplitude. As $l_\phi$ is limited mainly by electron-electron interaction at low temperatures \cite{Hansen2001}, it may be affected by DC bias voltage, but further investigation is required for a better understanding.



\subsection*{Additional data on the interplay of AB oscillations and Klein tunneling} 

\begin{figure}[h!]\centering
\includegraphics[draft=false,keepaspectratio=true,clip,width=1\linewidth]{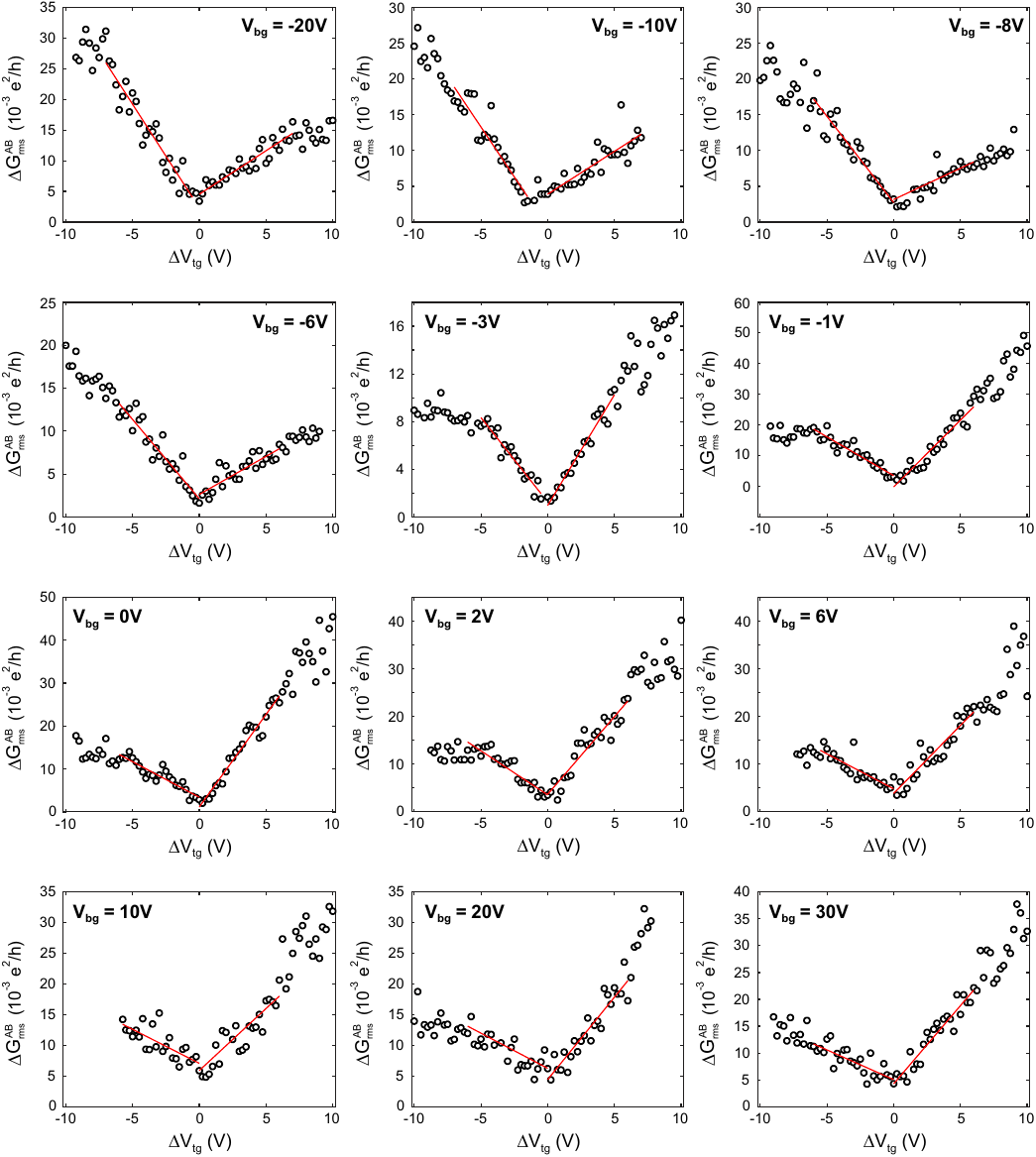}
\caption[figS10]{Summary of additional plots of AB amplitudes as function of $\Delta V_{\text{tg}}=V_{\text{tg}}-V_{\text{tg}}^0$ for various back gate voltages (see labels).}
\label{figS10}
\end{figure}

In Fig.~\ref{figS10} we present all other data sets of AB amplitude versus $\Delta V_{\text{tg}}$ for different $V_{\text{bg}}$ in order to investigate Klein tunneling by AB oscillations. Each panel displays the AB amplitude by the RMS value of $\Delta G$, $\Delta G_{\text{rms}}^{AB}$, as function of $V_{\text{tg}}$ for a specific $V_{\text{bg}}$. Together with the data shown in the main text $V_{\text{bg}}$ ranges from $-30$\,V to $30$\,V. The linear fits for the extraction of the amplitude asymmetry are included in the amplitude profiles . As described in main text we observe an asymmetry around the top-gate CNP. In vicinity of the back-gate CNP of the device the profiles are almost symmetric (see e.g. $V_{\text{bg}}=-3$\,V). The asymmetry switches at the  back-gate CNP and increases towards high back-gate voltages.
\newpage

\newpage
\section{Supplementary details on theory}

\subsection*{Theoretical model for the oscillatory conductance}

In this theoretical analysis, we model our device as a scattering region between two leads. We assume that these leads are identical and host a number of states, denoted by $N(E_F)$, which depends on the Fermi energy.
As our experiments are performed at very low temperature, we assume that $T=0$ in our theoretical model.
Taking into account both spin and valley degeneracy, we can then express the conductance of our device as~\cite{Buettiker1985}
\begin{equation}
  G(E_{\text{F}}) = \frac{I(V_{\text{b}})}{V_{\text{b}}} = 4 \frac{e^2}{h} \sum_{m,n}^{N(E_F)} T_{n,m}(E_F) ,
  \label{eq:I-V-zerobias}
\end{equation}
which is valid in the linear approximation in the bias voltage $V_b$. The factor $T_{n,m}(E_F)$ expresses the transmission probability from channel $m$ in the incoming lead to channel $n$ in the outgoing lead. Throughout this section, we consider positive Fermi energies $E_{\text{F}}$, i.e. we consider incoming electrons. Because of electron-hole symmetry, the corresponding results for holes can easily be derived. 

Our device consists of a ring, with a lower and and upper arm. In the lower arm, a top gate voltage $V_\text{tg}$ induces a potential energy $U(x,y)$, with maximum $U_0$.
A complete calculation of the transmission through the device takes into account direct transmission through the upper and lower arms of the ring, as well as paths that encircle the entire ring, either once or multiple times. For a ring with a single channel, all these contributions were added in Ref.~\cite{Gefen1984}, where the authors obtained an analytical result for the transmission. Their result is valid in the limit where the phase coherence length $l_\phi$ is much larger than the size of the ring. In our experiment, $l_\phi$ is however on the order of $\pi\overline{r}$, as was previously extracted from the data in Fig.~2, Figs.~\ref{figS12a} and~\ref{figS12ab}. Due to loss of coherence, the contribution to the Aharonov-Bohm oscillations of paths that encircle the ring $n$ times is suppressed by a factor $\exp(- 2\pi n \overline{r}/l_\phi)$, which is small in our case. These paths, which show up as higher harmonics in the Fourier spectrum, therefore play a much smaller role in the transmission. We thus believe that we can obtain a good estimate for the total transmission by taking only direct transmission through the upper and lower ring arms into account.


We can label the modes in the incoming lead by their transversal momentum $p_n$. Electrons in these modes are partially transmitted to the various modes in the upper and lower ring arms, and partially reflected. The electrons that are transmitted into the ring arms are subsequently partially transmitted to the modes in the outgoing lead. In principle, all incoming modes can couple to all outgoing modes through all different modes in the lead, as discussed in more detail in Ref.~\cite{Buettiker1985}. In this calculation, we will however assume that scattering between different modes only plays a minor role, and therefore neglect it. This means that an electron with transversal momentum $p_n$ is transmitted, with a certain probability, to a mode with transversal momentum $p_n$ in the ring arm, and subsequently transmitted to a mode with transversal momentum $p_n$ in the outgoing lead, again with a certain probability.
The idea behind this assumption is that $p_n$ is conserved across the np- and nn$'$-junctions that arise in the lower ring arm, because we can model these junctions by a one-dimensional potential $U(x)$, as we discuss later in somewhat more detail. This description seems reasonable because the junction width is much smaller than the ring circumference ($w_\mathrm{pn}\ll\pi \overline{r}$). Admittedly, this assumption is fairly crude, and means that we certainly miss different contributions. However, we believe that the description allows us to capture the main mechanism, while at the same time keeping the model analytically tractable.

The first consequence of this assumption is that we only consider the diagonal elements $T_{nn}$ in equation~(\ref{eq:I-V-zerobias}), i.e.,
\begin{equation}
  G(E_{\text{F}}) = 4 \frac{e^2}{h} \sum_{n}^{N(E_F)} T_{n,n}(E_F) .
  \label{eq:I-V-zerobias-diag}
\end{equation}
Note that we add the contributions of the various modes incoherently in this equation.
Furthermore, the assumption also implies that we can write $T_{n,n}$ as
\begin{align}
  T_{n,n} &= \left| t_\mathrm{up}(E_{\text{F}},p_n) + t_\mathrm{lo}(E_{\text{F}},U_0,p_{n}) \right|^2  \\
    &= \left| t_\mathrm{up}(E_{\text{F}},p_n) \right|^2 + \left| t_\mathrm{lo}(E_{\text{F}},U_0,p_{n}) \right|^2 + 2 \, \mathrm{Re}\left[ t_\mathrm{up}^*(E_{\text{F}},p_n) t_\mathrm{lo}(E_{\text{F}},U_0,p_{n}) \right] 
\end{align} 
where $t_\mathrm{up(lo)}(p_n)$ represents the tunneling through the upper (lower) ring arm as a function of $p_n$.
%
We can write the transmission coefficients through the arms as
\begin{align}
  t_\mathrm{up}(E_{\text{F}},p_{n}) &= |t_\mathrm{up}(E_{\text{F}},p_{n})| \exp\left( i \phi_\text{up,AB} + i \phi_\text{up,dyn}(E_F,p_n) \right), \\
  t_\mathrm{lo}(E_{\text{F}},U_0,p_{n}) &= |t_\mathrm{lo}(E_{\text{F}},U_0,p_{n})| \exp\left( i \phi_\text{lo,AB} + i \phi_\text{lo,dyn}(E_F,U_0,p_n) \right),
 \label{eq:t-expansion}
\end{align}
where we have adopted a semiclassical picture in order to write the phase as a sum of the (geometrical) Aharonov-Bohm phase and the dynamical phase.
Adding all different channels, we obtain
\begin{equation}
  G(E_{\text{F}}) = 4 \frac{e^2}{h} \sum_{n}^{N(E_F)} \left| t_\mathrm{up}(E_{\text{F}},p_n) \right|^2 + \left| t_\mathrm{lo}(E_{\text{F}},U_0,p_{n}) \right|^2 + 2 \, \mathrm{Re}\left[ t_\mathrm{up}^*(E_{\text{F}},p_n) t_\mathrm{lo}(E_{\text{F}},U_0,p_{n}) \right]  .
  \label{eq:G-added}
\end{equation}
As the magnetic field is weak, and mainly serves as a probe, we can assume that the absolute values of the tunneling amplitudes do not depend on it. This approximation implies that the first two terms in the above equation are independent of the magnetic field and do not contribute to the Aharonov-Bohm oscillations. The last term describes interference between electrons going through the different ring arms and thus describes the Aharonov-Bohm oscillations.
Note that one could, in principle, include the magnetic field in the tunneling amplitudes using the results from Ref.~\cite{Shytov2008}.



Using equation~(\ref{eq:t-expansion}), we can therefore write the oscillatory part of the conductance as~(\ref{eq:G-added}) as
\begin{equation}
  \Delta G^\mathrm{AB} = 8 \frac{e^2}{h} \mathrm{Re}\left[ \sum_{n}^{N(E_F)} |t_\mathrm{up}| e^{- i \phi_\text{up,AB}} e^{-i \phi_\text{up,dyn}} |t_\mathrm{lo}| e^{ i \phi_\text{lo,AB}} e^{i \phi_\text{lo,dyn}} \right].
\end{equation}
To a first approximation, the Aharonov-Bohm phase is independent of $p_n$, as the majority of the magnetic flux goes through the ring opening, and a much smaller part through the arms of the ring. This means that we can take the Aharonov-Bohm phase out of the summation and use Stokes' theorem to obtain
\begin{equation}
  \Delta G^\mathrm{AB} = 8 \frac{e^2}{h} \mathrm{Re}\left[ \exp\left(\frac{i e}{\hbar} \oint B \cdot \text{d} S \right) \sum_{n}^{N(E_F)} |t_\mathrm{up}| |t_\mathrm{lo}| e^{i \phi_\text{lo,dyn} -i \phi_\text{up,dyn}} \right].
\end{equation}
In order to determine the root mean square of the Aharonov-Bohm oscillations,
\begin{equation}
  \Delta G^\mathrm{AB}_{\mathrm{rms}} = 8 \frac{e^2}{h} \left\langle \left( \mathrm{Re}\left[ \exp\left(\frac{i e}{\hbar} \oint B \cdot \text{d} S \right) \sum_{n}^{N(E_F)} |t_\mathrm{up}| |t_\mathrm{lo}| e^{i \phi_\text{lo,dyn} -i \phi_\text{up,dyn}} \right] \right)^2 \right\rangle_B^{1/2} ,
\end{equation}
where $\langle \,\ldots\, \rangle_B$ denotes the averaging over the magnetic field, we use the identity
\begin{equation}
  (\mathrm{Re}[z])^2 = \frac{z^2}{4} + \frac{(z^*)^2}{4} + \frac{|z|^2}{2} .
\end{equation}
Since in our case $z$ contains an oscillating function, only the third term remains after the averaging. We thus obtain
\begin{equation}
  \Delta G^\mathrm{AB}_{\mathrm{rms}} = 4\sqrt{2} \, \frac{e^2}{h} \left| \sum_{n}^{N(E_F)} |t_\mathrm{up}(E_{\text{F}},p_{n})| |t_\mathrm{lo}(E_{\text{F}},U_0,p_{n})| e^{i [\phi_\text{lo,dyn}(E_{\text{F}},U_0,p_{n}) - \phi_\text{up,dyn}(E_{\text{F}},p_{n})]} \right|.
\end{equation}

We now assume that the dynamical phase only weakly depends on $p_n$. In the lowest order approximation, we can then neglect this dependence, and take the phase factor outside of the summation. Due to the absolute value, it subsequently disappears from the result. In the npn-regime this assumption can be partially justified, since the transmission is largest around normal incidence (zero transversal momentum), and decays rapidly with the angle of incidence, depending on the effective width $d_\text{pn}$, as we discuss in the following section. We also remark that this assumption leads to an overestimation of the transmission amplitudes, as the triangle inequality gives
\begin{multline} \label{eq:sum-triagle-inequality}
  \left| \sum_n^{N(E_F)} |t_\mathrm{up}(E_{\text{F}},p_{n})| |t_\mathrm{lo}(E_{\text{F}},U_0,p_{n})| e^{i[\phi_\text{lo,dyn}(E_{\text{F}},U_0,p_{n})- \phi_\text{up,dyn}(E_{\text{F}},p_{n})]} \right| 
  \\ 
  \leq \sum_n^{N(E_F)} |t_\mathrm{up}(E_{\text{F}},p_{n})| |t_\mathrm{lo}(E_{\text{F}},U_0,p_{n})| .
\end{multline}
Note that, according to our assumption, a change in the gate potential leads to a global phase shift of the Aharonov-Bohm oscillations, once the background has been subtracted. Previous experiments~\cite{Ford1989b} on gated Aharonov-Bohm rings in GaAs/AlGaAs heterostructures show that this system actually exhibits much richer behavior. In these experiments, the phase seemed to be ``pinned'' near $B=0$, as it did not change substantially between different top gate voltages. This pinning may have been a consequence of the Onsager relations, see also Refs.~\cite{Buettiker1985,Yacoby1996}.
In the experiment~\cite{Ford1989b}, amplitude modulations as a function of the top gate voltage $V_\text{tg}$, which controls the potential $U_0$, were also observed. However, contrary to our experimental observations, these modulations were periodic. 
This leads us to believe that, in our experiment, the change in the root mean square of the Aharonov-Bohm oscillations between different top gate voltages is mainly caused by the modification of the transmission coefficient, and only to a lesser extent by the modification of the interference, in agreement with our model.
Our approximation for $\Delta G^\mathrm{AB}_{\mathrm{rms}}$ therefore reads
\begin{multline} \label{eq:delta-G-rms}
  \Delta G^\mathrm{AB}_{\mathrm{rms}} = 4\sqrt{2} \, \frac{e^2}{h} N(E_F) |t_\mathrm{pr}| \;\;\; \mathrm{with} \\ |t_\mathrm{pr}(E_F,U_0)| = \frac{1}{N(E_F)}\sum_n^{N(E_F)} |t_\mathrm{up}(E_{\text{F}},p_{n})| |t_\mathrm{lo}(E_{\text{F}},U_0,p_{n})|,
\end{multline}
where we have introduced the average tunneling amplitude $|t_\mathrm{pr}|$, which equals the product of the transmission probabilities through the upper and lower ring arms, averaged over all modes in the lead. We note that its value lies between 0 and 1.

Our next step is to obtain a formula for the average tunneling amplitude $|t_\mathrm{pr}|$. As both the leads and the ring arms are quite wide, we invoke the continuum approximation to perform this computation. We thus assume that all parts of the system host a continuum of states. Within this limit, the quantized momentum $p_n$ becomes the continuous transversal momentum $p_y$. As we already discussed before, we model the np- and nn$'$-junctions, which arise in the lower ring arm, by a one-dimensional potential $U(x)$, since the junction width is much smaller than the ring circumference ($w_\mathrm{pn}\ll\pi \overline{r}$). This implies that $p_y$ is conserved across the junction interface.
When we take the continuum limit, we therefore convert the sum over the various modes in the definition of $|t_\text{pr}|$ into an integral over $p_y$ and incorporate the distribution of the transversal momenta through a continuous distribution function $f(p_y)$. To ensure that the transmission lies between zero and one, we normalize by the integral of this distribution. This leads to
\begin{equation} \label{eq:t-prod-av-cont}
  |t_\mathrm{pr}(E_F,U_0)| = \frac{\int f(p_y) |t_\mathrm{up}(E_{\text{F}},p_{y})| |t_\mathrm{lo}(E_{\text{F}},U_0,p_{y})| \, \mathrm{d}p_y}{\int f(p_y) \, \mathrm{d}p_y} .
\end{equation}
Studies of billiard models~\cite{Beenakker89,Milovanovic13,Milovanovic15,Reijnders17,Reijnders17b} show that the initial distribution $f(p_y)$ can usually be well approximated by a uniform distribution.
Since the upper arm of the device is not gated, we set $|t_\mathrm{up}(E_{\text{F}},p_{y})| = 1$, which means that we neglect scattering at the inlet of the ring.

Since we have invoked the continuum approximation, the charge carriers obey the classical dispersion relation~\cite{Reijnders13,Tudorovskiy12} $(U(x)-E_{\text{F}})^2 = p_x^2(x) + p_y^2$, which shows that a classically forbidden region appears for $-v_{\text{F}} |p_y| < U(x) -E_{\text{F}} < v_{\text{F}}|p_y|$. 
When the maximum $U_0$ of the potential $U(x)$ satisfies $U_0<E_{\text{F}}$, we have two nn$'$-junctions in the lower ring arm. However, by virtue of the dispersion relation, we only find classically allowed electron states between these two junctions for incoming states that have a transversal momentum which satisfies $v_{\text{F}} |p_y| \leq E_{\text{F}} - U_0$. Thus, more states become available inside the lower ring arm when we decrease the potential $U_0$, as the inequality is satisfied for more values of $p_y$, which, by virtue of equation~(\ref{eq:t-prod-av-cont}), leads to an increase in $|t_\mathrm{pr}|$ and therefore to an increase in the amplitude~(\ref{eq:delta-G-rms}) of the Aharonov-Bohm oscillations.
When there are classically allowed electron states in the lower ring arm, we set the tunneling amplitude $t_{nn'}(E_{\text{F}},p_y)$ to unity. When these states are absent, we set the tunneling amplitude to zero. Although, in the latter case, charge carriers can theoretically tunnel through the region between the two junctions, the corresponding amplitude is exponentially small~\cite{Reijnders13} and can be neglected. Likewise, we neglect the exponentially small reflection for the nn$'$-junction~\cite{Reijnders13}.


Similar considerations hold for the case $U_0>E_{\text{F}}$, which leads to an np- and a pn-junction in the lower ring arm.
There are classically allowed hole states in the lower ring arm when the transversal momentum of the incoming electron state satisfies $v_F |p_y| \leq U_0 - E_{\text{F}}$. This inequality is satisfied for more values of $p_y$ when $U_0$ increases, similar to the previous case. However, this time the tunneling amplitude $t_{np}(E_{\text{F}},p_y)$ is smaller than one, as the electrons have to tunnel through a classically forbidden region~\cite{Reijnders13,Tudorovskiy12}. By virtue of equation~(\ref{eq:t-prod-av-cont}), this implies that the slope of $\Delta G^\mathrm{AB}_{\mathrm{rms}}$ is smaller for $U_0>E_{\text{F}}$ than for $U_0<E_{\text{F}}$.
These considerations qualitatively explain the shape of the curves in Figs.~3a-d of the main text and Fig.~\ref{figS10}.


\subsection*{Derivation of the tunneling amplitude}

In order to obtain a more quantitative comparison between theory and experiment, we compute the average tunneling amplitude $|t_\mathrm{pr}|$ explicitly. We focus on the case of an npn-junction, noting that similar considerations hold for the nn$'$n-junction.

We first note that we can neglect multiple reflections within the junction. Since both the mean free path $l_m$ and the phase coherence length $l_\phi$ of the charge carriers are smaller than or on the order of $\pi\overline{r}$, the contribution of multiple reflections to the transmission amplitude is suppressed.
Within a semiclassical framework~\cite{Reijnders13}, we can then write down the absolute value of the tunneling amplitude through the lower ring arm as
\begin{equation}
  |t_\mathrm{lo}(E_{\text{F}},U_0,p_y)| = |t_\mathrm{np}(E_{\text{F}},U_0,p_y)|  |t_\mathrm{pn}(E_{\text{F}},U_0,p_y)| .
  \label{eq:t-down-factorized}
\end{equation}
Since we consider the absolute value of the tunneling amplitude, we do not have to compute its phase, which contains the classical action of the particle along its trajectory in the lower ring arm.

Previously, we assumed that scattering between different modes only plays a minor role, and therefore neglected it. In the continuum approximation, this enables the factorization~(\ref{eq:t-down-factorized}). We nevertheless note that this is a fairly crude approximation, since the magnetic length is about $115$~nm for a magnetic field of $50$~mT. Within a continuum semiclassical picture, the transversal momentum $p_y$ may therefore change somewhat within the lower (and similarly in the upper) ring arm. We nevertheless resort to this approximation, as computing the variation of $p_y$ requires very large additional efforts. Furthermore, we neglect edge scattering, which is strongly suppressed for sliding electrons with zigzag edges~\cite{Dugaev13}. Note that generic edges behave fairly similar to zigzag edges~\cite{Akhmerov08} and in particular do not lead to valley mixing.


We can obtain an analytic expression for $|t_\mathrm{lo}(E_{\text{F}},U_0,p_y)|$ by considering a linear potential $U(x)$, i.e., $U(x) = U_0 \; x/ w_{\text{pn}}$ for $0<x<w_{\text{pn}}$ and $U(x)$ is constant outside of this regime.
The tunneling amplitude $t_\mathrm{np}(E_{\text{F}},U_0,p_y)$ is then given by~\cite{Cheianov2006,Tudorovskiy12}
\begin{equation}
  t_\mathrm{np}(E_{\text{F}},U_0,p_y) = \exp\left(-\frac{\pi}{2} \frac{w_\mathrm{np}}{\hbar v_{\text{F}} U_0} v_{\text{F}}^2 p_y^2\right)
    = \exp\left(- \frac{\pi^2 d_\mathrm{np}}{U_0 E_{\text{F}}} v_{\text{F}}^2 p_y^2\right) ,
  \label{eq:t-np-linear}
\end{equation}
where we have introduced the dimensionless effective junction width $d_\mathrm{np} = w_\mathrm{np} E_{\text{F}}/(2\pi\hbar v_{\text{F}})$.
The same expression holds for $t_\mathrm{pn}$ when we consider a linear pn-junction.
Using equations~(\ref{eq:t-prod-av-cont}), (\ref{eq:t-down-factorized}) and~(\ref{eq:t-np-linear}) and assuming that $w_\mathrm{np}=w_\mathrm{pn}$ ($d_\mathrm{np}=d_\mathrm{pn}$), we then obtain
\begin{multline}
  |t_\mathrm{pr}(E_F,U_0)| 
    = \; \frac{v_F}{2E_{\text{F}}} \int_{-(U_0-E_{\text{F}})/v_{\text{F}}}^{(U_0-E_{\text{F}})/v_{\text{F}}} \exp\left( - \frac{2\pi^2 d_\mathrm{pn}}{U_0 E_{\text{F}}} v_{\text{F}}^2 \; p_y^2 \right) \mathrm{d}p_y \\
    = \;\sqrt{\frac{U_0}{2\pi^2 d_\mathrm{pn} E_{\text{F}}}} \int_0^{\sqrt{\frac{2\pi^2 d_\mathrm{pn}}{U_0 E_{\text{F}}}}(U_0-E_{\text{F}})} e^{-y^2} \mathrm{d} y .
\end{multline}
Note that we integrate from $-(U_0-E_{\text{F}})/v_{\text{F}}$ to $(U_0-E_{\text{F}})/v_{\text{F}}$, since the tunneling amplitude is zero for values of $|p_y|$ outside of this regime because we are no longer dealing with an npn-junction.
Using the definition of the so-called error function, $\mathrm{Erf}(x)=\frac{2}{\sqrt{\pi}} \int_0^x \exp(-y^2) \mathrm{d}y$, we can rewrite this expression as
\begin{equation}
  |t_\mathrm{pr}(E_F,U_0)| 
    = \sqrt{\frac{U_0}{8 \pi d_\mathrm{pn} E_{\text{F}}}} \; \mathrm{Erf}\left( \sqrt{\frac{2\pi^2 d_\mathrm{pn}}{U_0 E_{\text{F}}}} (U_0-E_{\text{F}}) \, \right) .
    \label{eq:t-lo-linear-erf}
\end{equation}
Note that this expression is only valid for $E_{\text{F}}\leq U_0 \leq 2E_{\text{F}}$, as, within our approximation, no new states in the lower ring arm become available beyond $U_0=2 E_{\text{F}}$.

Expression~(\ref{eq:t-lo-linear-erf}) shows that $|t_\mathrm{pr}(E_F,U_0)|$ increases by two mechanisms when we increase the potential $U_0$. First of all, the error function increases. From a physical point of view, this corresponds to new modes that become available within the lower ring arm. Second, the prefactor increases, which corresponds to increased transmission through the available modes.
We can now identify two regimes in our model. For small $U_0$, we can expand the error function to first order in its argument to obtain $|t_\mathrm{pr}(E_{\text{F}},U_0)| = (U_0-E_{\text{F}})/E_{\text{F}}$. Hence, the combination of both mechanisms leads to a linear increase of the average tunneling amplitude $|t_\mathrm{pr}|$. At large values of $U_0$, the error function has saturated and the second mechanism dominates: the tunneling amplitude increases like a square root.

\begin{figure}[h]\centering
\includegraphics[draft=false,width=1\linewidth]{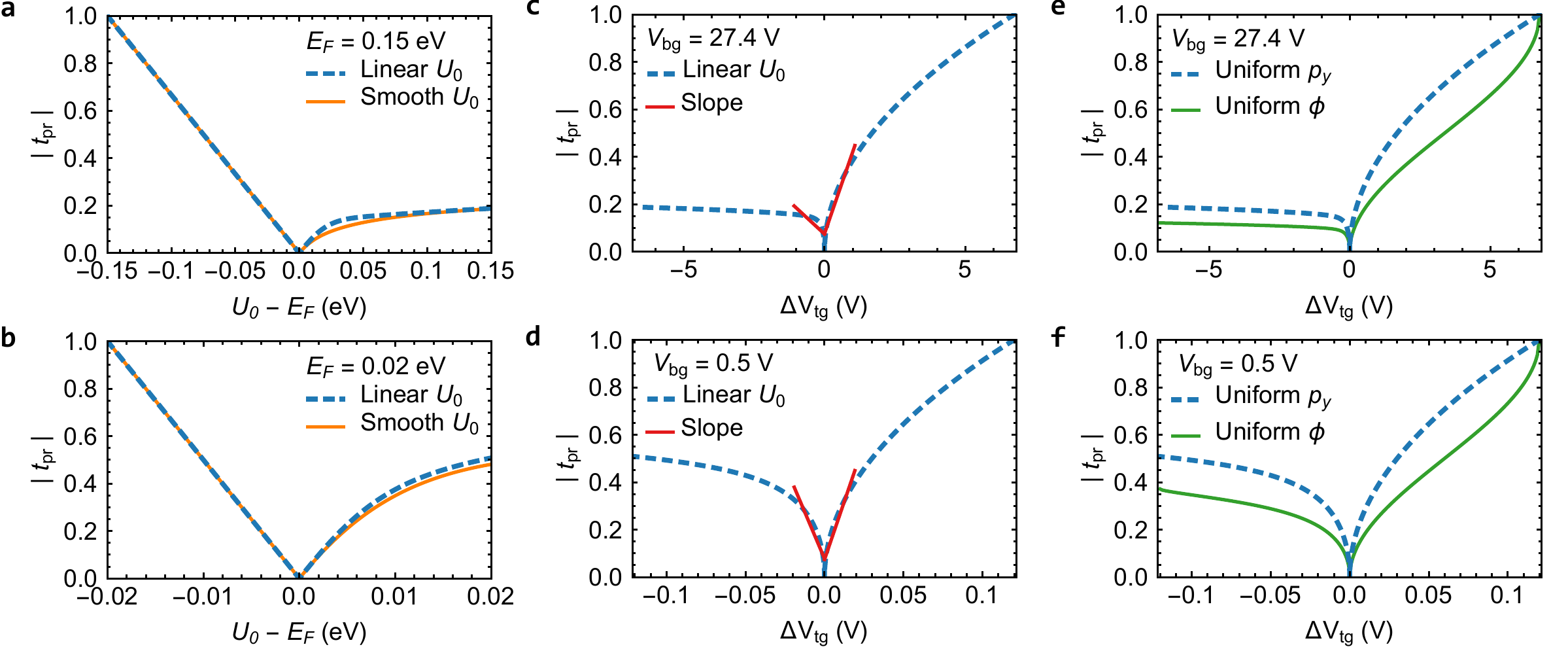}
\caption{Theoretical prediction for the averaged tunneling amplitude $|t_\mathrm{pr}(E_{\text{F}},U_0)|$. (a), (b) Comparison between a linear potential and the smooth potential~(\ref{eq:U-smooth}), for two different Fermi energies. (c), (d) Determination of the slope for the back gate voltages corresponding to the Fermi energies in panels (a) and (b). (e), (f) Effect of the distribution $f(p_y)$ for a linear potential. We consider a distribution uniform in emission angles $\phi$ and a distribution uniform in transversal momenta $p_y$.
}
\label{fig:t-E-V-dep}
\end{figure}

Since $\mathrm{Erf}(\sqrt{2})\approx 0.95$, we may estimate that the transition between these two regimes lies close to
\begin{equation}
  U_0 = E_{\text{F}} \left( 1 + \frac{1 + \sqrt{1 + (2 \pi)^2 d_\mathrm{pn}}}{2\pi^2 d_\mathrm{pn}} \right) .
  \label{eq:U0-transition}
\end{equation}
Crucially, this factor strongly depends on the effective junction width $d_\mathrm{pn}$. At small $d_\mathrm{pn}$, we mainly observe the first regime where $|t_\mathrm{pr}|$ increases linearly. This leads to a symmetric situation where the npn- and nn$'$n-junctions display similar behavior, see Fig.~\ref{fig:t-E-V-dep}(b).
On the other hand, for large $d_\mathrm{pn}$ we mainly observe the second regime, which leads to a much slower increase in $|t_\mathrm{pr}|$, see Fig.~\ref{fig:t-E-V-dep}(a).

In order to obtain results for more realistic potentials $U(x)$, we can use the semiclassical approximation for the tunneling amplitude. Within this scheme, $t_\mathrm{np}(E_{\text{F}},U_0,p_y)$ is given by~\cite{Reijnders13,Tudorovskiy12,Shytov2008,Cheianov2006}
\begin{equation}
  t_\mathrm{np}(E_{\text{F}},U_0,p_y) = e^{-d_\mathrm{np} K_\mathrm{np}}, \;\;   \mathrm{with} \;\;\;
  d_\mathrm{np} K_\mathrm{np} = \frac{1}{\hbar v_{\text{F}}} \int_{x_-}^{x_+} \sqrt{v_{\text{F}}^2 p_y^2 - (U(x) - E_{\text{F}})^2} \, \mathrm{d}x ,
\end{equation}
where $K_\mathrm{np}$ is the classical action in the classically forbidden region and $x_{\pm}$ are the classical turning points, which satisfy $U_0(x_\pm) = E_{\text{F}} \pm v_{\text{F}} |p_y|$. Note that this approximation correctly reproduces the exact result~(\ref{eq:t-np-linear}) for a linear potential.
As shown in Figs.~\ref{fig:t-E-V-dep}(a) and \ref{fig:t-E-V-dep}(b), the smooth potential increase
\begin{equation}
  U(x) = \frac{U_0}{2} \left(1 +  \tanh\left( \frac{2 x}{w_\mathrm{pn}}-1 \right) \right) ,
  \label{eq:U-smooth}
\end{equation}
which has the same slope at $x=w_\mathrm{pn}/2$ as the linear potential, leads to qualitatively similar results for the averaged tunneling amplitude $|t_\mathrm{pr}(E_{\text{F}},U_0)|$. 


\subsection*{Derivation of the normalized difference of the slopes}

To compare our theoretical predictions with the experimental observations, we extract the slopes of $\Delta G_{\mathrm{rms}}^{\mathrm{AB}}$. 
Within our theoretical model, this slope is directly proportional to the slope of $|t_\mathrm{pr}|$, see equation~(\ref{eq:delta-G-rms}), where the proportionality depends on the Fermi energy $E_F$.
However, our theoretical model predicts that large increases in the transmission only occur for $0 < U_0 < 2 E_{\text{F}}$.  
Figs.~\ref{fig:t-E-V-dep}(c) and~\ref{fig:t-E-V-dep}(d) show that this leads to a fitting regime $\Delta V_{\text{tg}}$ that strongly depends on $V_{\text{bg}}$, i.e. on $E_{\text{F}}$, in contrast to the experimental observations shown in Fig.~\ref{figS10}.
This difference is due to the many simplifications that we have made throughout the model. In particular, the upper and lower ring arms contain less modes than the leads, which implies that states can also be filled outside of the regime $0 < U_0 < 2 E_{\text{F}}$, leading to an increase in the amplitude of the Aharonov-Bohm oscillations. 
However, we may assume that our simplifications affect the transmission on both sides of the charge neutrality point in a similar way, which means that the prediction for their ratio is much more accurate. Rather than comparing the individual slopes with the experimental data, we therefore focus on the normalized difference $A_\text{npn}$ of the slopes.



Our next step is to carefully think about an appropriate interval $\Delta V_\mathrm{tg}$ that we can use to determine the slopes,
as we cannot use the experimental interval.
As we discussed in the previous section, our model exhibits two regimes. The transition point between these two regimes is roughly given by equation~(\ref{eq:U0-transition}) and crucially depends on $d_\mathrm{pn}$.
Since large increases in the transmission occur for $U_0$ between $0$ and $2 E_{\text{F}}$, we parametrize our fitting interval as $(1-\alpha)E_{\text{F}} \leq U_0 \leq E_{\text{F}}$ for the nn$'$n regime and $E_{\text{F}} \leq U_0 \leq (1+\alpha) E_{\text{F}}$ for the npn regime.
If we choose $\alpha$ too small, we overestimate the influence of the first regime. On the other hand, if we choose it too large, the influence of the second regime is overestimated. We therefore believe that $\alpha$ should ideally lie somewhere around $0.5$.
After choosing $\alpha$, we convert these energy intervals into intervals for the top gate voltage $\Delta V_\mathrm{tg}$.

\begin{figure}[t]\centering
\includegraphics[draft=false,width=0.75\linewidth]{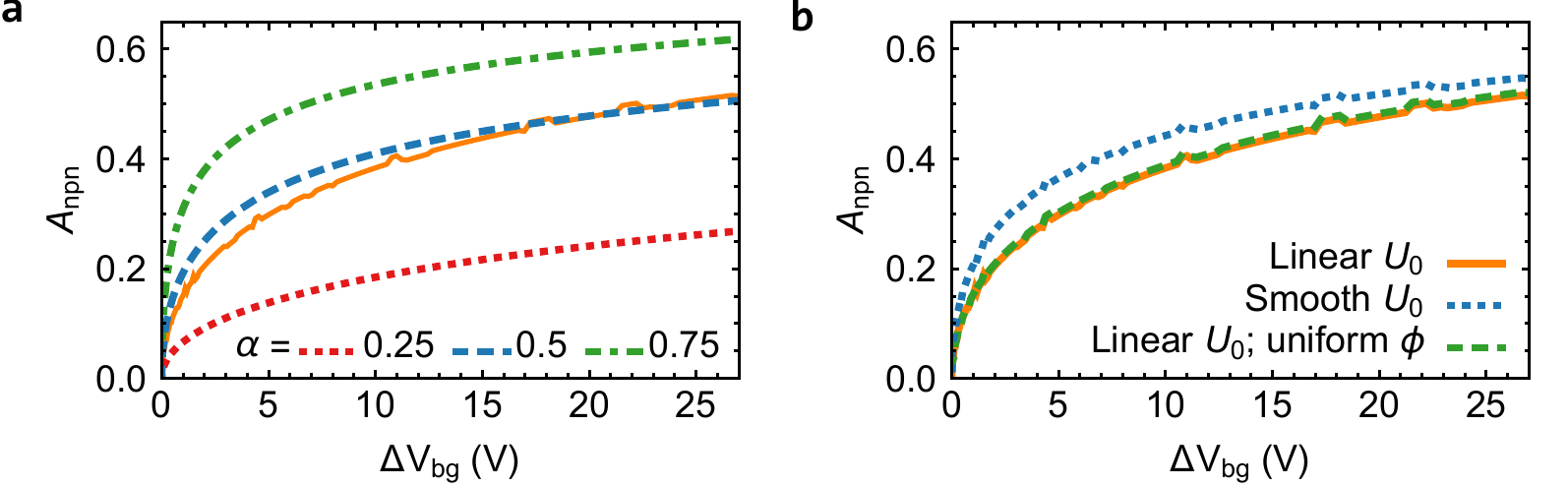}
\caption{(a) Normalized difference of the slopes obtained from formula~(\ref{eq:A-np-nn-alpha}) for different values of $\alpha$ (dotted, dashed and dashed-dotted lines). We also show the result obtained by determining the slopes of $|t_\mathrm{pr}|$ for a linear potential using $\alpha=0.4$ (solid line).
(b) Normalized difference of the slopes obtained by determining the slopes. We compare the linear potential (solid line) with the smooth potential~(\ref{eq:U-smooth}) (dotted line). We also consider a linear potential with a uniform distribution of the emission angles $\phi$ (dashed line).
}
\label{fig:A-dep}
\end{figure}

We can obtain an analytic result for the normalized difference of the slopes $A_\text{npn}$ by approximating the slopes by the difference quotient, that is, by setting
\begin{equation}
  a_\text{npn} = 4\sqrt{2} \, \frac{e^2}{h} N(E_F) \frac{|t_\mathrm{pr}(E_{\text{F}}, (1+ \alpha)E_{\text{F}})| - |t_\mathrm{pr}(E_{\text{F}}, E_{\text{F}})|}{|\Delta V_\mathrm{tg}((1+ \alpha)E_{\text{F}} ) - \Delta V_\mathrm{tg}( E_{\text{F}} )|} .
\end{equation}
Equation~(\ref{eq:t-lo-linear-erf}) shows that the transmission vanishes at the charge neutrality point, in other words that $|t_\mathrm{pr}(E_{\text{F}}, E_{\text{F}})| = 0$. Since we also have $\Delta V_\mathrm{tg}( E_{\text{F}} ) = 0$, we arrive at
\begin{equation}
  A_\text{npn} = \frac{|a_\text{nn'n}|-|a_\text{npn}|}{|a_\text{nn'n}|+|a_\text{npn}|} = \frac{1 - \chi}{1 + \chi}, 
  \qquad \chi = \frac{|t_\mathrm{pr}(E_{\text{F}}, (1+ \alpha)E_{\text{F}})| }{|\Delta V_\mathrm{tg}((1+ \alpha)E_{\text{F}})| } \frac{|\Delta V_\mathrm{tg}((1 - \alpha)E_{\text{F}})| }{|t_\mathrm{pr}(E_{\text{F}}, (1 - \alpha)E_{\text{F}})| } .
\end{equation}
Combining equation~(\ref{eq:t-lo-linear-erf}) with the observations that $|\Delta V_\mathrm{tg}((1 + \alpha)E_{\text{F}})| = |\Delta V_\mathrm{tg}((1 - \alpha)E_{\text{F}})|$ and that 
$|t_\mathrm{pr}(E_{\text{F}}, (1 - \alpha)E_{\text{F}})| = \alpha$,
we finally obtain
\begin{align}
  A_\text{npn} &= \frac{1 - \frac{1}{2\alpha}\sqrt{\frac{1+\alpha}{2 \pi d_\mathrm{pn}}} \; \mathrm{Erf}\left( \pi \alpha \sqrt{\frac{2 d_\mathrm{pn}}{1+\alpha}} \right)}{1 + \frac{1}{2 \alpha}\sqrt{\frac{1+\alpha}{2 \pi d_\mathrm{pn}}} \; \mathrm{Erf}\left( \pi \alpha \sqrt{\frac{2 d_\mathrm{pn}}{1+\alpha}} \right)} ,
%
%
\nonumber\\
A_\text{pnp} &= \frac{|a_\text{pp'p}|-|a_\text{pnp}|}{|a_\text{pp'p}|+|a_\text{pnp}|} = A_\text{npn}.
  \label{eq:A-np-nn-alpha}
\end{align}
Note that all prefactors have canceled in this equation, and the normalized difference of the slopes only depends on the transmission coefficients.
Setting $\alpha=1/2$ in formula~(\ref{eq:A-np-nn-alpha}), we arrive at equation~(3) of the main text.
In Fig.~\ref{fig:A-dep}(a), we plot our result~(\ref{eq:A-np-nn-alpha}) for three different values of $\alpha$.
Importantly, all three curves predict a similar dependence of $A_\text{npn}$ on $\Delta V_\text{bg}$. 
However, the attained values depend on the value of $\alpha$. 
When $\alpha=0.25$, we attach greater importance to the first one of the regimes discussed in the previous subsection, leading to relatively small values of $A_\text{npn}$. On the other hand, we attach greater importance to the second regime when $\alpha=0.75$, leading to much larger values of $A_\text{npn}$. From a theoretical point of view, setting $\alpha=0.5$ seems to be a good compromise.

We can compare our equation~(\ref{eq:A-np-nn-alpha}) with the result obtained by determining the slopes of the transmission~(\ref{eq:t-lo-linear-erf}). When we use the fitting intervals discussed above with $\alpha=0.4$, we observe that the result nicely coincides with our equation~(\ref{eq:A-np-nn-alpha}) with $\alpha=0.5$, as shown in Fig.~\ref{fig:A-dep}. We performed the curve fitting for the graphs of $|t_\mathrm{pr}|$ versus $\Delta V_{tg}$, as shown in Figs.~\ref{fig:t-E-V-dep}(c) and~\ref{fig:t-E-V-dep}(d), in order to be consistent with the procedure for the experimental data. However, performing the fitting for the graphs of $|t_\mathrm{pr}|$ versus $U_0$ does not substantially change the result.

In Fig.~\ref{fig:A-dep}(b), we compare the normalized difference of the slopes extracted for a linear potential and for the smooth potential~(\ref{eq:U-smooth}). We observe that a smooth potential leads to exactly the same shape of the curve, although slightly higher values of $A_\text{npn}$ are attained. We also show what happens when we use a uniform angular distribution for $f(p_y)$ in equation~(\ref{eq:t-prod-av-cont}), instead of a distribution that is uniform in the transversal momentum $p_y$. Although the transmission curves are significantly altered, as shown in Figs.~\ref{fig:t-E-V-dep}(e) and~\ref{fig:t-E-V-dep}(f), the normalized difference of the slopes remains the same. Hence, we conclude that the distribution function $f(p_y)$ has only a minor influence on $A_\text{npn}$ (and $A_\text{pnp}$).

\bibliographystyle{apsrev4-1}

\bibliography{diss}